\documentclass{aa}  
\usepackage{natbib}
\bibpunct{(}{)}{;}{a}{}{,} 

\usepackage[T1]{fontenc}
\usepackage{graphicx}
\usepackage{amssymb}
\usepackage{subcaption}
\usepackage{cleveref}
\usepackage{booktabs}

\usepackage{epstopdf}
\usepackage{gensymb}
\usepackage{siunitx}
\usepackage{tikz}
\usepackage{mathtools}
\usepackage{soul}
\usepackage{commath}
\usepackage{afterpage}
\usepackage{placeins}
\usepackage{diagbox}
\usepackage{makecell}
\usepackage{tabularx}

\begin{document}

   \title{System equivalent flux density of a low-frequency polarimetric phased array interferometer}


   \author{A. T. Sutinjo
          \inst{1}
           \and
           D. C. X. Ung
          \inst{1}
          \and
          M. Sokolowski
          \inst{1}
           \and
          M. Kovaleva
          \inst{1}
           \and
          S. McSweeney
          \inst{1}
          }

   \institute{International Centre for Radio Astronomy Research (ICRAR), Curtin University, 6102 Australia\\
              \email{adrian.sutinjo@curtin.edu.au}
             }

   \date{
   Revision 1 \today
   }

 
  \abstract
  {} 
   {This paper extends the treatment of system equivalent flux density (SEFD) in~\citet{Sutinjo_AA2021} (Paper~I) to interferometric phased array telescopes. The objective is to develop an SEFD formula involving only the most fundamental assumptions and one that is readily applicable to phased array interferometer radio observations. Then, we aimed at comparing the resultant SEFD expression against the often-used root-mean-square (RMS) SEFD approximation, $\text{SEFD}_{I}^{rms}=\frac{1}{2} \sqrt{\text{SEFD}_{XX}^2 + \text{SEFD}_{YY}^2}$ to study the inaccuracy of the $\text{SEFD}_{I}^{rms}$.}
   {We take into account all mutual coupling and noise coupling within an array environment (intra-array coupling). This intra-array noise coupling is included in the SEFD expression through the realized noise resistance of the array, which accounts for the system noise.  
   No assumption is made regarding the polarization (or lack thereof) of the sky nor the orthogonality of the antenna elements. The fundamental noise assumption is that, in phasor representation, the real and imaginary components of a given noise source are  independent and equally distributed (iid) with zero mean. Noise sources that are mutually correlated and non-iid among themselves are allowed, provided the real and imaginary components of each noise source are iid. The system noise is uncorrelated between array entities separated by a baseline distance, which in the case of the Murchison Widefield Array (MWA) is typically tens of wavelengths or greater. By comparing the resulting SEFD formula to the $\text{SEFD}_{I}^{rms}$ approximation, we proved that $\text{SEFD}_{I}^{rms}$ always underestimates the SEFD, which leads to an overestimation of array sensitivity.}
   {We present the resulting SEFD formula that is generalized for the phased array, but has a similar form to the earlier result in Paper~I. Here, the physical meaning of the antenna lengths and the equivalent noise resistances have been generalized such that they are also valid in the array environment. The simulated SEFD was validated using MWA observation of a Hydra-A radio galaxy at 154.88~MHz. The observed $\text{SEFD}_{XX}$ and $\text{SEFD}_I$ are on average higher by 9\% and 4\% respectively, while the observed $\text{SEFD}_{YY}$ is lower by 4\% compared to simulated values for all pixels within the $-12$~dB beam width. The simulated and observed SEFD errors due to the RMS SEFD approximation are nearly identical, with mean difference of images of virtually 0\%. This result suggests that the derived SEFD expression, as well as the simulation approach, is correct and may be applied to any pointing. As a result, this method permits identification of phased array telescope pointing angles where the RMS approximation underestimates SEFD (overestimates sensitivity). For example, for Hydra-A observation with beam pointing (Az, ZA) = (81\degree, 46\degree), the underestimation in SEFD calculation using the RMS expression is 7\% within $-3$~dB beam width but increases to 23\% within $-12$~dB beam width. At 199.68~MHz, for the simulated MWA pointing at (Az, ZA) = (45\degree, 56.96\degree), the underestimation reached 29\% within $-3$~dB beam width and 36\% within $-12$~dB beam width. This underestimation due to RMS SEFD approximation at two different pointing angles and frequencies was expected and is consistent with the proof.}
 {}

\keywords{Instrumentation: interferometers--Techniques: polarimetric--Methods: analytical--
Methods: data analysis--Methods: numerical--
Methods: observational--Telescopes}

\maketitle
%
\section{Introduction}
\label{sec:intro}

Our previous work in~\citet{Sutinjo_AA2021} (hereafter referred to as Paper~I) discussed a formulation for the sensitivity in terms of System Equivalent Flux Density (SEFD) for a polarimetric interferometer that consists of dual polarized antennas. In that work, we provided an example using a dual-polarized MWA dipole element embedded in an array. In this paper, we extend and generalize the SEFD formulation to polarimetric phased arrays interferometers. This is an important generalization as it is directly applicable to low-frequency interferometric phased array telescopes in operation such as the  Murchison Widefield Array (MWA)~\citep{2013PASA...30....7T} and LOFAR~\citep{2013A&A...556A...2V}, as well as future Low Frequency Square Kilometre Array (SKA-Low)~\citep{8105424, 7928622}. In particular regarding the SKA-Low, a clear conceptual understanding of array sensitivity, how it varies over telescope pointing angles, and how to calculate it are crucial for validation against SKA sensitivity requirements~\citep{SKA1L1req}.   

The reasons for using SEFD as the valid figure of merit (FoM) for a polarimetric radio interferometer as opposed to antenna effective area on system temperature ($A_e/T_{sys}$) was thoroughly reviewed in Paper I. However for convenience, we mention the key ideas here. The primary reason is that antenna effective area, $A_e$, is a number that is defined as matched to the polarization of the incident wave, which is not known in advance for a polarimeter. In contrast, the concept of equivalent flux density is not constrained to the polarization state and it can be readily equated to the system noise. The work in Paper~I allowed us to demonstrate that the often used conversion between $A_e/T_{sys}$ and SEFD is only an approximation that is valid in certain cases in which the Jones matrix is diagonal or anti-diagonal. This is further generalized in our current paper (see  Sec.~\ref{sec:RMSunderestimates}), where we show that the SEFD approximation is correct only for row vectors in the Jones matrix (see Eq.~\eqref{eqn:J1}) that are orthogonal. When the row vectors are not orthogonal, the root-mean-square (RMS) SEFD approximation always underestimates the true SEFD. 

Furthermore, in our work on this subject, we make an explicit connection to observational radio interferometry which is demonstrated by comparison to radio images. We also calculated and conducted a careful review of second-order noise statistics that form the basis for the SEFD formula. These are the main differences between our work and existing work in radio astronomy phased array sensitivity literature, for example~\citet{Warnick_6018280, Ellingson_5722983, Tokarsky_7987810, 7293140}. However, we acknowledge that there are aspects of our calculations that benefit from the pre-existing collective knowledge in the community, in particular regarding the computation of array response to an incident wave and array noise temperature as discussed in Sec.~\ref{sec:Jones_mat} and Sec.~\ref{sec:Sys_noise}. 

\begin{figure}[t]
\begin{center}
\noindent
  \includegraphics[width=3.5in]{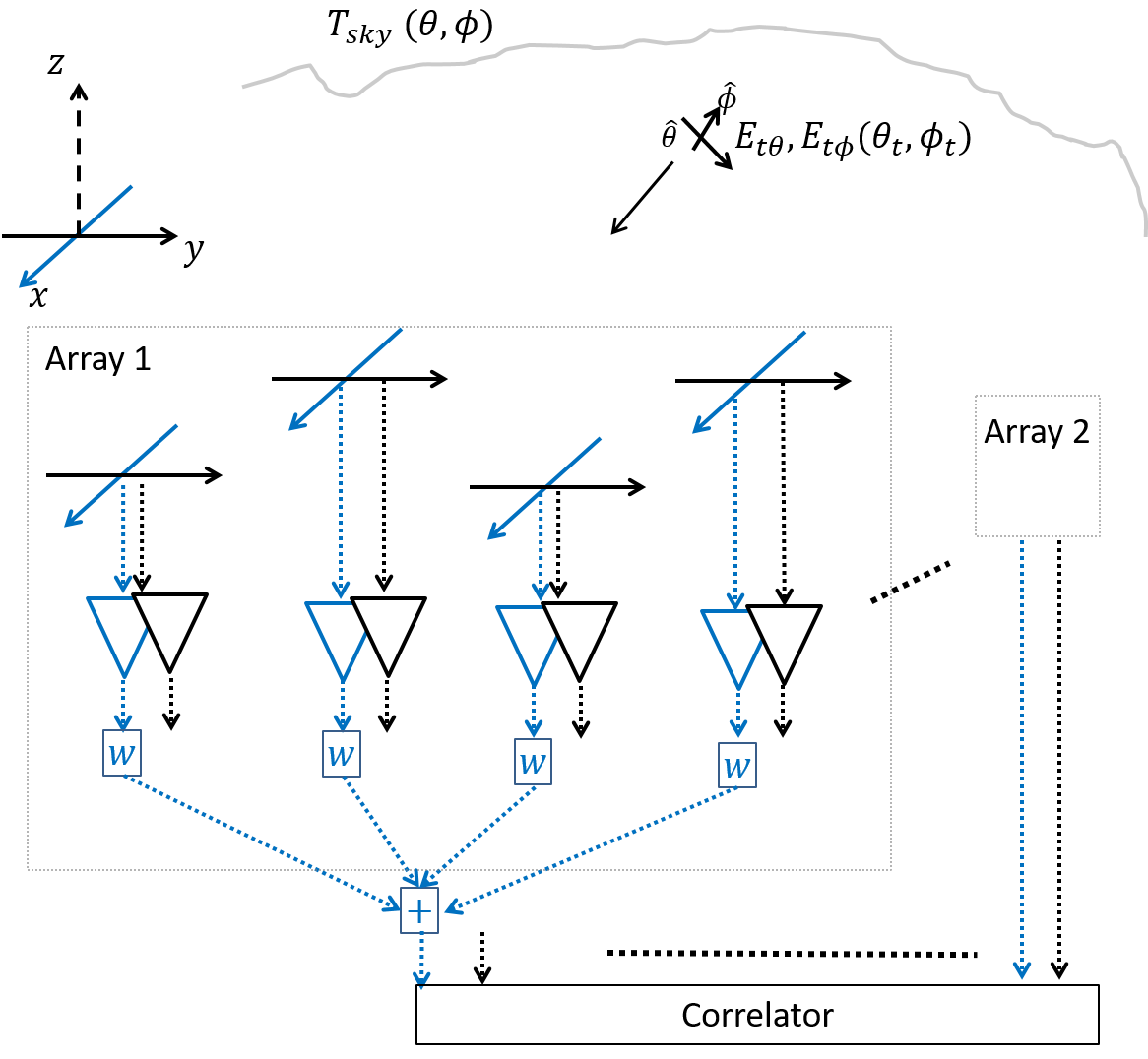}
\caption{Arrays of dual-polarized low-frequency antennas observing the sky. The antennas are at fixed positions above the ground and the $z$-axis is up. The signal from each element is amplified by an LNA, weighted, and summed. The output of each summer is directed into the correlator. The $y$-directed elements (black) follow the same signal flow, but not shown for simplicity. \textcolor{black}{A four-element array is shown as an example.}}
\label{fig:arrays}
\end{center}
\end{figure}

The polarimetric phased array under consideration has N dual-polarized elements where each element is connected to a low noise amplifier (LNA), as is typical in practice. The LNA outputs are connected to the phased array weights and subsequently summed to produce the array output as shown in Fig.~\ref{fig:arrays}. We consider the array response to an incident electric field from a target direction, $\mathbf{e}_t$. The voltages that represent the response of Array~1 are
\begin{eqnarray}
\mathbf{v}_1|_t&=&\mathbf{J}\mathbf{e}_t, \nonumber \\
\left[ \begin{array}{c}
V_{X1}|_t \\
V_{Y1}|_t 
\end{array} \right]
&=&
\left[ \begin{array}{cc}
l_{X\theta} & l_{X\phi} \\
l_{Y\theta} & l_{Y\phi}
\end{array} \right]
\left[ \begin{array}{c}
E_{t\theta} \\
E_{t\phi} 
\end{array} \right],
\label{eqn:J1}
\end{eqnarray} 
where $V_{X1}, V_{Y1}$ are the voltages measured by the $X, Y$ arrays, respectively, and $\mathbf{J}$ is the Jones matrix of the array; the elements of the Jones matrix are antenna lengths with units of meter that represent the response of the array to each polarization basis of the electric field, $E_{t\theta},E_{t\phi}$, with units of \si{\volt\per\metre}. There is a similar  equation corresponding to Array~ 2. For antenna arrays of an identical design, it is reasonable to assume the same Jones matrix as Array~ 1.

Equation~\eqref{eqn:J1} is applicable to an array of antennas. We consider an array of dual-polarized antennas with outputs that are weighted and summed, then connected to a correlator as depicted in Fig.~\ref{fig:arrays}, thereby forming an interferometer. The antennas receive noise from the sky which includes a partially polarized target source $\bar{E}_t$ and a background with noise temperature distribution given by $T_{sky}$. For brevity, only four elements are shown per array. The extension to N elements is immediately evident. The LNAs, represented by the triangle gain blocks, produce their own noise which is scattered by the array and picked up by all elements in the array. Similarly, the incident sky signal arriving at the array undergoes the same scattering and coupling process. Therefore, the combined noise voltages seen at the LNA inputs are mutually correlated. We consider these noise sources and the interactions thereof to extend the SEFD formula to the interferometric and polarimetric phased array telescope. 

This paper is organized as follows. The SEFD derivation and calculation are presented in Sec.~\ref{sec:SEFD_calc}. An example SEFD calculation procedure for the MWA is demonstrated in Sec.~\ref{sec:simulation}. 
The SEFD calculation is validated by MWA observations described in Sec.~\ref{sec:obs} with comparison and discussion given in Sec.~\ref{sec:compare}. Concluding remarks are summarized in Sec.~\ref{sec:concl}. The appendix presents a detailed review of fundamental assumptions and statistical  calculations that justify the SEFD formula presented in Sec.~\ref{sec:SEFD_calc}.

\section{SEFD of a polarimetric phased array interferometer}
\label{sec:SEFD_calc}
\subsection{Jones matrix of a dual-polarized array}
\label{sec:Jones_mat}
The calculations in Paper~I are reusable for an array but need reinterpretation and adaptation which we now discuss. The first key modification is to take the direction-dependent antenna lengths as that which are seen in the array environment, that is where each antenna element is connected to the input of an LNA. We call this quantity the antenna ``realized length'' \citep{Ung_9040892, Ung_MPhil2020} which is different from the open circuit antenna effective length in Paper~I. 

\begin{figure}[t]
\begin{center}
\noindent
  \includegraphics[width=3in]{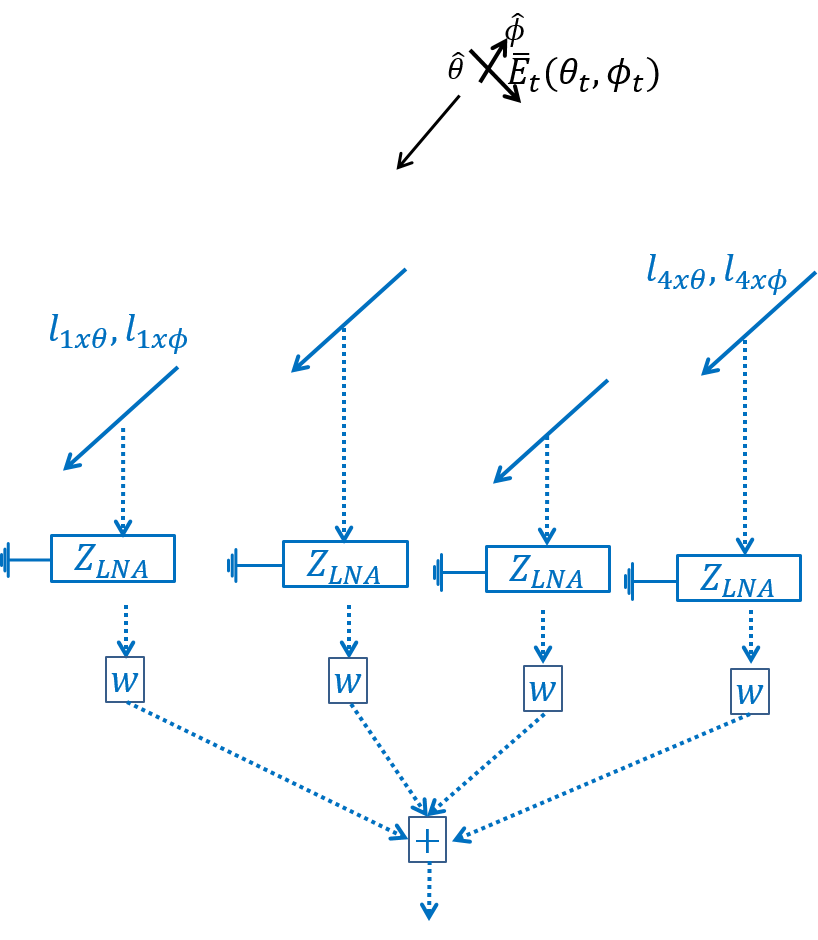}
\caption{Antenna length calculation for an array, where $l_{1x\theta}$ is the antenna length associated with $\hat{\theta}$-polarized incident field for $x$-directed antenna in the embedded element number 1 in the array shown, similarly with $l_{ix\theta}, l_{iy\theta}$, where $i$ is the embedded element number. \textcolor{black}{For brevity we consider a four-element array $i=1,\cdots,4$; the extension to $i=1,\cdots,N$ presents no complication.}}
\label{fig:ant_len_cal}
\end{center}
\end{figure}

The realized length has the advantage of having a clear and physical interpretation in an array environment depicted in Fig.~\ref{fig:ant_len_cal}. It is obtained by summing the embedded element realized lengths to form an overall equivalent realized length for the array. For example, for the X array,
\begin{eqnarray}
l_{X\theta}=\mathbf{w}_{x}^T\mathbf{l}_{x\theta}, \nonumber \\
l_{X\phi}=\mathbf{w}_{x}^T\mathbf{l}_{x\phi},
\label{eqn:array_lX}
\end{eqnarray} 
where $\mathbf{w}_{x}^T=[w_{1x},...,w_{4x}]$ is the vector of weights and $\mathbf{l}_{x\theta}^T=[l_{1x\theta},...,l_{4x\theta}]$ is a vector containing the embedded element realized lengths, and similarly with $\mathbf{l}_{x\phi}$. The quantities $l_{1x\theta}$ and $l_{1x\phi}$ are obtained for a dual-polarized embedded element (number 1) with all other elements in the array terminated with the LNA input impedance, $Z_{LNA}$, as shown in Fig.~\ref{fig:ant_len_cal}. The embedded antenna realized length can be obtained through full-wave electromagnetic simulation~\citep{Ung_9040892} or measurement. Similarly, for the Y array
\begin{eqnarray}
l_{Y\theta}=\mathbf{w}_{y}^T\mathbf{l}_{y\theta}, \nonumber \\
l_{Y\phi}=\mathbf{w}_{y}^T\mathbf{l}_{y\phi},
\label{eqn:array_lY}
\end{eqnarray} 
where $\mathbf{w}_{y}^T=[w_{1y},...,w_{4y}]$ is the vector of weights for the Y array, which may differ from $\mathbf{w}_{x}$. We note that Eq.~\eqref{eqn:array_lX} and Eq.~\eqref{eqn:array_lY} apply to the antenna realized length as we emphasized earlier, as opposed to the open-circuit length. 
We assume LNAs of an identical design in this paper, and  hence the LNA voltage gains are identical and need not be explicitly shown in Eq.~\eqref{eqn:J1}. 

Given the foregoing discussion regarding the realized length, then for a phased array, the voltages on the left hand side of Eq.~\eqref{eqn:J1} represent those at the outputs of the summer shown in Fig.~\ref{fig:arrays}. Therefore, the Jones matrix, $\mathbf{J}$, translates the incident field to the voltages at the outputs of the summer. This is similarly the case for Array 2 in Fig.~\ref{fig:arrays}. The entries of $\mathbf{J}$ represent the realized lengths of the array (for X or Y component) to an electric field basis ($\hat{\theta}$ or $\hat{\phi}$) such that each entry is a linear combination of the realized lengths of the embedded elements in the array. The Jones matrix then, is
\begin{eqnarray}
\mathbf{J}=\left[ \begin{array}{cc}
l_{X\theta} & l_{X\phi} \\
l_{Y\theta} & l_{Y\phi}
\end{array} \right]= 
\left[ \begin{array}{cc}
\mathbf{w}_x^T \mathbf{l}_{x\theta} & \mathbf{w}_x^T \mathbf{l}_{x\phi} \\
\mathbf{w}_y^T \mathbf{l}_{y\theta} & \mathbf{w}_y^T \mathbf{l}_{y\phi}
\end{array} \right].
\label{eqn:J_array}
\end{eqnarray} 
We adopt the uppercase ``X'' or ``Y'' to denote the overall array response and lower case ``$x$'' or ``$y$'' for the embedded elements in the array.

\subsection{SEFD expression}
\label{sec:SEFD exp}
The interferometric polarimeter estimates the outer product of the incident electric field in the target direction, which from Paper~I is given by
\begin{eqnarray}
\tilde{\mathbf{e}}\tilde{\mathbf{e}}^H=\mathbf{J}^{-1}\mathbf{v}_1\mathbf{v}_2^H\mathbf{J}^{-H},
\label{eqn:pol}
\end{eqnarray}
where $\mathbf{J}$ is defined by Eq.~\eqref{eqn:J_array} which represents the response of the array. The SEFD is proportional to the standard deviation of the flux estimate  
\begin{eqnarray}
\text{SEFD}_I=\frac{\text{SDev}(\tilde{I})}{\eta_0}=\frac{\text{SDev}\left[(\tilde{\mathbf{e}}\tilde{\mathbf{e}}^H)_{1,1}+(\tilde{\mathbf{e}}\tilde{\mathbf{e}}^H)_{2,2}\right]}{\eta_0},
\label{eqn:SEFD_STDev}
\end{eqnarray}
where $\__{1,1}, \__{2,2}$ indicate the diagonal entries of the matrix in Eq.~\eqref{eqn:pol} (hence the subscript $_I$ that refers to Stokes I) and $\eta_0 \approx 120\pi~\Omega$ is the free space impedance. What is needed is the expression for the standard deviations (SDev) for the phased array interferometer, which we develop next.

\subsection{Array system noise}
\label{sec:Sys_noise}
We now compute the standard deviation of the electric field estimates as expressed in Eq.~\eqref{eqn:SEFD_STDev} due to the system noise. As shown in Paper~I, we expect to express the standard deviation in Eq.~\eqref{eqn:SEFD_STDev} in terms of the array realized lengths and the system noise temperatures, $T_{sys}$. Therefore, we need expressions for the mean square voltages for the X and Y arrays due to the system noise. 

\begin{figure}[t]
\begin{center}
\noindent
  \includegraphics[width=3.5in]{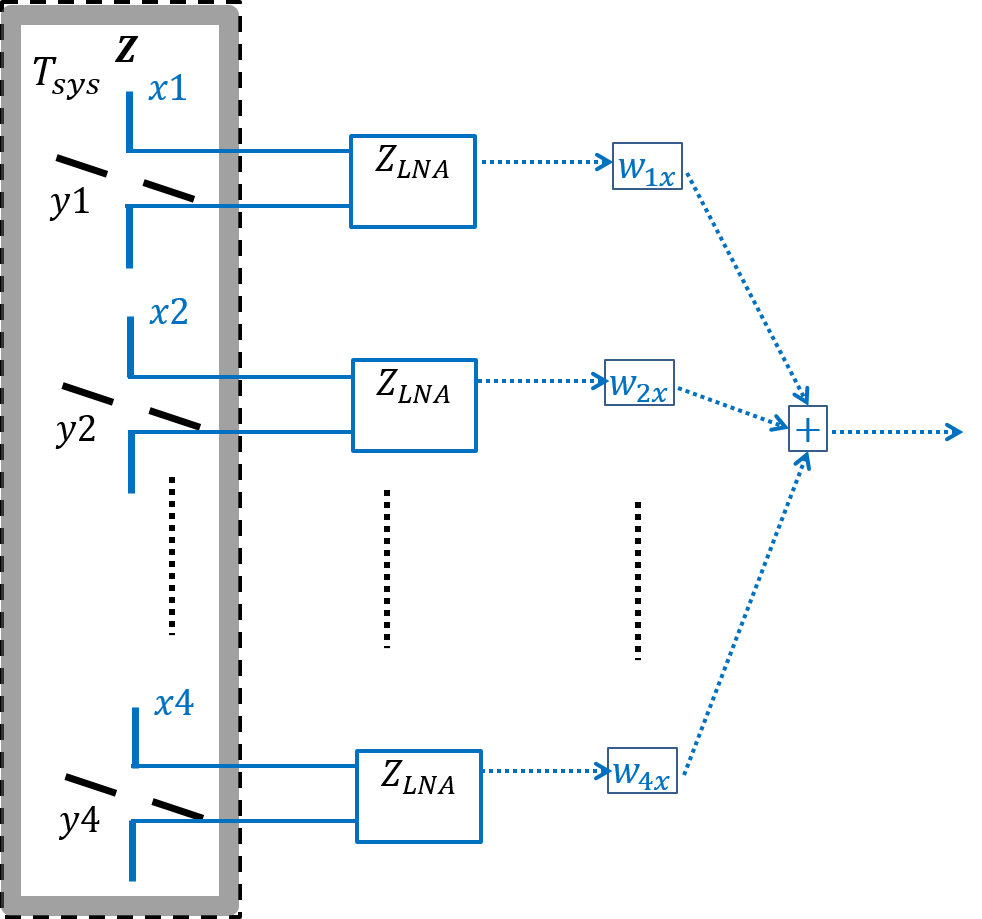}
\caption{Diagrammatic representation of system noise. The array is surrounded by a homogeneous blackbody radiator at temperature $T_{sys}$ that represents the diffuse sky noise and the LNA noise. The $y$ elements are similarly connected to a separate weight and summing circuit that are not shown for brevity.}
\label{fig:Tsys}
\end{center}
\end{figure}

Fig.~\ref{fig:Tsys} depicts an array illuminated by an equivalent homogeneous black body environment at temperature $T_{sys}$. The value of $T_{sys}$ is that which equals the noise due to the actual diffuse background sky under observation, $T_{sky}$, plus the noise of the LNAs ($T_{rcv})$ in the array environment, and the ohmic loss which is a function of the radiation efficiency ($\eta_{rad}$) of the array. The $T_{sys}$ value for a phased array can be obtained through computation or measurements or a combination thereof using well-documented procedures. For example, computation for $T_{rcv}$ of an array was discussed in the literature~\citep{Warnick_5062509, Belo_7079488,  warnick_maaskant_ivashina_davidson_jeffs_2018, Ung_8739942, Ung_9040892}. The computation of $T_{sys}$ for the MWA was demonstrated in~\citet{Ung_9040892, Ung_MPhil2020} and was validated by observation. The $T_{sys}$ value could also be obtained through measurements by carefully calibrating the array gain via known hot and cold sources, for example see~\citet{chippendale_hayman_hay_2014}. 

The $T_{sys}$ for the X and Y arrays are generally different; this is similarly noted in~Paper~I for a dual-polarized antenna element. This is expected because for the same sky under observation, the X and Y arrays produce different responses on the sky. Moreover, the $x$ elements and the $y$ elements interact differently depending on the array configuration and weights. Therefore, we anticipate assigning different $T_{sysX}$ and $T_{sysY}$ to the respective arrays, which we discuss next.

An equivalent homogeneous sky at $T_{sys}$ gives rise to partially correlated noise voltages at the antenna ports in the array.  A convenient way to quantify this is to start with the case where each antenna port is terminated with an open circuit, that is $Z_{LNA}\rightarrow\infty$. In this case, the mutual coherence is known from the generalized Nyquist theorem~\citep{doi:10.1063/1.1722048, Hillbrand_1084200}
\begin{eqnarray}
\mathbf{C}_o|_{T_{sys}}=\left<\mathbf{v}_o\mathbf{v}_o^H\right>|_{T_{sys}}=4kT_{sys}\Delta f\Re [\mathbf{Z}], 
\label{eqn:Re} 
\label{eqn:twiss}
\end{eqnarray}
where $\Delta f$ is the noise bandwidth and $\Re [\mathbf{Z}]$ is the real part of the antenna impedance matrix, which is a real and symmetric matrix. We choose the port numbering convention such that 
\begin{eqnarray}
\Re [\mathbf{Z}]=\Re
\left[\begin{array}{cc} 
\mathbf{Z}_{xx} & \mathbf{Z}_{xy} \\
\mathbf{Z}_{yx}  & \mathbf{Z}_{yy}
\end{array} \right],
\label{eqn:ReZ}
\end{eqnarray}
that is the $x_1,\cdots,x_4$ embedded elements ports are numbered 1 to 4 and $y_1,\cdots,y_4$ embedded elements ports are numbered 5 to 8, such that the open circuit voltage vector is $\mathbf{v}_o^T=[V_{ox1},\cdots,V_{ox4},V_{oy1},\cdots,V_{oy4}]$. The open circuit voltage vector is related to the voltage seen at the LNA inputs through a transformation matrix~\citep{Warnick_5062509}, which we call $\mathbf{T}$, such that
\begin{eqnarray}
\mathbf{v}=Z_{LNA}\left[Z_{LNA}\mathbf{I}+\mathbf{Z}\right]^{-1}\mathbf{v}_{o}=\mathbf{T}\mathbf{v}_{o}.
\label{eqn:v_voc}
\end{eqnarray}
Therefore, the coherence matrix of the voltages seen at the inputs of the LNAs loaded with $Z_{LNA}$ is
\begin{eqnarray}
\mathbf{T}\mathbf{C}_o|_{T_{sys}}\mathbf{T}^H&=&\mathbf{T}\left<\mathbf{v}_o\mathbf{v}_o^H\right>|_{T_{sys}}\mathbf{T}^H \nonumber \\
&=&4kT_{sys}\Delta f\mathbf{T}\Re [\mathbf{Z}]\mathbf{T}^H,
\label{eqn:coh_v_LNA}
\end{eqnarray} 
which is a Hermitian matrix, since $(\mathbf{T}\Re [\mathbf{Z}]\mathbf{T}^H)^H=\mathbf{T}\Re [\mathbf{Z}]\mathbf{T}^H$. 

The sought after quantities are the mean square voltages after weighting and summing 
\begin{eqnarray}
\left<|V_{X}|^2\right>&=&[\mathbf{w}_{x}^T,\mathbf{0}_4^T] \mathbf{T}\mathbf{C}_o|_{T_{sysX}}\mathbf{T}^H
\left[\begin{array}{c} 
\mathbf{w}_{x}^* \\ \mathbf{0}_4
\end{array} \right] \nonumber \\
&=&4kT_{sysX}\Delta f\mathbf{w}_{x0}^T\mathbf{T}\Re [\mathbf{Z}]\mathbf{T}^H\mathbf{w}_{x0}^*, \nonumber \\ 
\left<|V_{Y}|^2\right>&=&[\mathbf{0}_4^T,\mathbf{w}_{y}^T] \mathbf{T}\mathbf{C}_o|_{T_{sysY}}\mathbf{T}^H
\left[\begin{array}{c} 
\mathbf{0}_4 \\ \mathbf{w}_{y}^* 
\end{array} \right] \nonumber \\
&=&4kT_{sysY}\Delta f\mathbf{w}_{y0}^T\mathbf{T}\Re [\mathbf{Z}]\mathbf{T}^H\mathbf{w}_{y0}^*,  
\label{eqn:ms_after_weight}
\end{eqnarray}
where $\mathbf{0}_4$ is a 4-long column vector of zeros; $\mathbf{w}_{x0}, \mathbf{w}_{y0}$ reflect the fact that the $x$ and $y$ antennas are summed separately; $T_{sysX}$ and $T_{sysY}$ are distinguished as discussed previously; $k$ is the Boltzmann constant and $\Delta f$ is the noise bandwidth. We know that Eq.~\eqref{eqn:ms_after_weight} is real because the eigenvalues of a Hermitian matrix are real.
For brevity in the SEFD formula, we adopt the following shorthand notations
\begin{eqnarray}
R_X=\mathbf{w}_{x0}^T\mathbf{T}\Re [\mathbf{Z}]\mathbf{T}^H\mathbf{w}_{x0}^*, \nonumber \\
R_Y=\mathbf{w}_{y0}^T\mathbf{T}\Re [\mathbf{Z}]\mathbf{T}^H\mathbf{w}_{y0}^*,
\label{eqn:R_noise}
\end{eqnarray}
which we can think of as the realized array noise resistances, with units of \si{\ohm}, representing the noise voltages at the outputs of the summers. A detailed explanation of this quantity and comparison to current literature can be found in Appendix \ref{apendx:links}. We note that Eq.~\eqref{eqn:ms_after_weight} transforms $T_{sysX}, T_{sysY}$ from a quantity external to the array to the mean square voltages after weighting and summing by the array. Therefore, the output of the summer is the reference plane at which the SEFD is to be calculated, as we show next.

\subsection{$\text{SEFD}_I$ formulation}
\label{sec:SEFD calc}
We assume all antenna arrays in question are of an identical design and the coupling between arrays is negligible due to the inter-array distance of tens of wavelength or larger. The intra-array coupling within each array is accounted for through the process described in Sec.~\ref{sec:Jones_mat} and Sec.~\ref{sec:Sys_noise}. This is the same level of assumptions as~Paper~I. The main difference between Paper~I and this work is that the sensitivity formula is now extended and generalized to the phased array. Following from~Paper~I, the Stokes~I estimate is $\tilde{I}=(\tilde{\mathbf{e}}\tilde{\mathbf{e}}^H)_{1,1}+(\tilde{\mathbf{e}}\tilde{\mathbf{e}}^H)_{2,2}$ from Eq.~\eqref{eqn:pol} and Eq.~\eqref{eqn:SEFD_STDev}. Expanding the matrix equation, we can write
\begin{eqnarray}
|D|^2\tilde{I}&=&X_1X_2^*\left(|l_{Y\phi}|^2+|l_{Y\theta}|^2\right) \nonumber \\
&+&Y_1Y_2^*\left(|l_{X\phi}|^2+|l_{X\theta}|^2\right) \nonumber \\
&-&X_1Y_2^*\left(l_{X\phi}^*l_{Y\phi}+l_{X\theta}^*l_{Y\theta}\right) \nonumber \\
&-&Y_1X_2^*\left(l_{X\phi}l_{Y\phi}^*+l_{X\theta}l_{Y\theta}^*\right),
\label{eqn:Isimp}
\end{eqnarray}
where  $D=l_{X\theta}l_{Y\phi}-l_{X\phi}l_{Y\theta}$ is the determinant of the array Jones matrix; the leading $V$'s have been suppressed for brevity, for example $X_1$ is a complex random variable that refers to $V_{X1}$. The components of the antenna realized lengths (which are direction-dependent complex scalars) now refer to that of the array as discussed in Sec.~\ref{sec:Jones_mat}. Following the statistical calculation described in Appendix \ref{apendx:stat},
\begin{eqnarray}
\frac{|D|^4\text{Var}(\tilde{I})}{{(4k\Delta   f)^2}}
&=&\mathbf{t}_R^T \mathbb{L}\mathbf{t}_R,
\label{eqn:var_I_simp_array}
\end{eqnarray}
where 
\begin{eqnarray}
\mathbf{t}_R=\left[ \begin{array}{c}
T_{sysX}R_X \\
T_{sysY}R_Y 
\end{array} \right],
\label{eqn:tr}
\end{eqnarray}
is a column vector and the matrix
\begin{eqnarray}
\mathbb{L}=\left[ \begin{array}{cc}
\norm{\mathbf{l}_Y}^4 & |l_{X\phi}^*l_{Y\phi}+l_{X\theta}^*l_{Y\theta}|^2 \\
|l_{X\phi}^*l_{Y\phi}+l_{X\theta}^*l_{Y\theta}|^2 & \norm{\mathbf{l}_X}^4
\end{array} \right],
\label{eqn:Lcal}
\end{eqnarray}
where the vector norms are $\norm{\mathbf{l}_X}^2=|l_{X\theta}|^2+|l_{X\phi}|^2$ and similarly with $\norm{\mathbf{l}_Y}^2$. We note in Eq.~\eqref{eqn:var_I_simp_array} that $R_X$ and $R_Y$ differ and generally cannot be factored out. Again, taking $\sqrt{\text{Var}(\tilde{I})}$ from Eq.~\eqref{eqn:var_I_simp_array} and dividing by the free space impedance, $\eta_0$, produces the desired formula
\begin{eqnarray}
\text{SEFD}_{I}=\frac{\text{SDev}(\tilde{I})}{\eta_0}=\frac{4 k\Delta f}{\eta_0}\frac{\sqrt{\mathbf{t}_R^T \mathbb{L}\mathbf{t}_R}}{|D|^2},
\label{eqn:SEFD_I_res_array}
\end{eqnarray}
with units \si{\watt\per\metre\squared}. If $\mathrm{SEFD}_I$ is stated in \si{\watt\per\metre\squared\per\hertz}, which is often preferred in radio astronomy, then we remove $\Delta f$ from the right hand side.

\subsection{RMS approximation always underestimates SEFD}
\label{sec:RMSunderestimates}
As mentioned in Sec.~\ref{sec:intro}, we now demonstrate the claim that the very commonly used RMS approximation in radio astronomy, for example see~\citep{1999ASPC..180..171W},  always underestimates the true SEFD. The RMS approximation is given by
\begin{eqnarray}
\text{SEFD}_I^{rms}= \frac{1}{2}\sqrt{\text{SEFD}_{XX}^2+\text{SEFD}_{YY}^2},
\label{eqn:SEFD_rms}
\end{eqnarray}
where $\text{SEFD}_{XX}, \text{SEFD}_{YY}$ assume an unpolarized source as discussed in Paper~I. For the notations and approach discussed in this paper, this approximation can be written as 
\begin{eqnarray}
\text{SEFD}_I^{rms}&=&\frac{4 k}{\eta_0}\sqrt{\frac{T_{sysX}^2R_X^2}{\norm{\mathbf{l}_X}^4}+\frac{T_{sysY}^2R_Y^2}{\norm{\mathbf{l}_Y}^4}}\nonumber\\
&=&\frac{4 k}{\eta_0}\sqrt{\mathbf{t}_R^T \mathbf{D}_{rms}\mathbf{t}_R},
\label{eqn:SEFD_rms_approx}
\end{eqnarray}
as pointed out in Paper~I. In addition, Appendix~\ref{apendx:RMS2L} shows that Eq.~\eqref{eqn:SEFD_rms_approx} is derivable using $A/T$. The diagonal matrix $\mathbf{D}_{rms}=\text{diag}\left[\norm{\mathbf{l}_{X}}^{-4}, \norm{\mathbf{l}_{Y}}^{-4}\right]$ is different from $\mathbb{L}/|D|^4$ in Eq.~\eqref{eqn:SEFD_I_res_array}. It can be shown that
\begin{eqnarray}
\mathbf{t}_R^T\frac{\mathbb{L}}{|D|^4}\mathbf{t}_R\geq \mathbf{t}_R^T\mathbf{D}_{rms}\mathbf{t}_R. 
\label{eqn:to_prove}
\end{eqnarray}

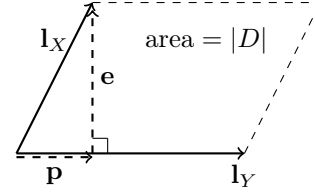
\begin{figure}
    \centering
    \begin{tikzpicture}
   \draw [thick, ->] (0,0) -- (3,0); 
\node at (3,-0.3) {$\mathbf{l}_Y$};
\draw [thick, ->] (0,0) -- (1,2);
\node at (0.5,1.5) {$\mathbf{l}_X$};
\draw [dashed, thick, ->] (0,-0.05) -- (1,-0.05);
\node at (0.5,-0.3) {$\mathbf{p}$};
\draw [dashed] (1,2) -- (4,2);
\draw [dashed] (3,0) -- (4,2);
\draw [dashed, thick, ->] (1,0) -- (1,1.9);
\node at (1.2,1.0) {$\mathbf{e}$};
\draw [-] (1.2,0.0) -- (1.2,0.2);
\draw [-] (1.2,0.2) -- (1.0,0.2);
\node at (2.5,1.5) {$\text{area}=|D|$};
\%draw [thick, <->] (0,2) -- (0,0) -- (2,0);
\%node at (1,1) {yes};
   \end{tikzpicture}
    \caption{Depiction of the vectors, their projections, and the area of the parallelogram.}
    \label{fig:e_pic}
\end{figure}

The reason for this is explained as follows. It can be shown through the Gram-Schmidt orthogonalization steps or the related QR factorization~\citep[see Ch.~4]{Strang_ILA2016} of the Jones matrix, that the absolute value of the determinant can be written as
\begin{eqnarray}
|D|=\norm{\mathbf{l}_Y}\norm{\mathbf{e}},
\label{eqn:error_proof}
\end{eqnarray}
where $\mathbf{e}$ is the projection vector of $\mathbf{l}_X$ onto a line which is perpendicular to $\mathbf{l}_Y$, as shown in Fig.~\ref{fig:e_pic} 
\begin{eqnarray}
\mathbf{e}=\mathbf{l}_X-\mathbf{p},
\label{eqn:error}
\end{eqnarray}
where $\mathbf{p}= (\mathbf{l}_Y^H\mathbf{l}_X)\mathbf{l}_Y/\norm{\mathbf{l}_Y}^2$,
is the orthogonal projection vector of $\mathbf{l}_X$ onto $\mathbf{l}_Y$~\citep[see Ch.~4]{Strang_ILA2016}, such that $\mathbf{e}$ and $\mathbf{p}$ are orthogonal, $\mathbf{e}\perp \mathbf{p}$.  Eq.~\eqref{eqn:error_proof} is easily verified by substitution of $\mathbf{l}_X=[l_{X\theta},l_{X\phi}]^T$ and $\mathbf{l}_Y=[l_{Y\theta},l_{Y\phi}]^T$ into Eq.~\eqref{eqn:error} and simplifying the vector algebra. There is also an insightful geometric interpretation of Eq.~\eqref{eqn:error_proof}: $|D|$ is the area formed by the parallelogram whose parallel sides are represented by vectors $\mathbf{l}_X$ and $\mathbf{l}_Y$ as shown in Fig.~\ref{fig:e_pic}; this is known from the volume property of determinants~\citep[see Ch.~5]{Strang_ILA2016} and is useful for understanding Eq.~\eqref{eqn:ineq_are_C2} discussed next.

Since $\norm{\mathbf{e}}\leq\norm{\mathbf{l}_X}$, then
\begin{eqnarray}
|D|^2\leq\norm{\mathbf{l}_{X}}^2\norm{\mathbf{l}_{Y}}^2.
\label{eqn:ineq_are_C2}
\end{eqnarray}
 The equality is fulfilled when the vectors are orthogonal, $\mathbf{l}_{Y} \perp \mathbf{l}_{X}$, which also implies $\mathbf{p}=\mathbf{0}$, $\mathbf{e}=\mathbf{l}_{X}$, and  $\mathbf{l}_{X}^H\mathbf{l}_{Y}=0$. This is fully expected given the above-mentioned geometric interpretation and proves Eq.~\eqref{eqn:to_prove}. To see this more clearly, we write 
\begin{eqnarray}
\frac{\mathbb{L}}{|D|^4}=\frac{1}{\norm{\mathbf{l}_X-\mathbf{p}}^4\norm{\mathbf{l}_{Y}}^4}\left[ \begin{array}{cc}
\norm{\mathbf{l}_{Y}}^4 & |\mathbf{l}_X^{H}\mathbf{l}_Y|^2 \\
|\mathbf{l}_X^{H}\mathbf{l}_Y|^2 & \norm{\mathbf{l}_{X}}^4
\end{array} \right],
\label{eqn:tLt_C2}
\end{eqnarray}
and 
\begin{eqnarray}
\mathbf{D}_{rms}=\frac{1}{\norm{\mathbf{l}_X}^4\norm{\mathbf{l}_{Y}}^4}\left[ \begin{array}{cc}
\norm{\mathbf{l}_{Y}}^4 & 0\\
0 & \norm{\mathbf{l}_{X}}^4
\end{array} \right].
\label{eqn:Drms_mat}
\end{eqnarray}
We note that the denominator on the right hand side of Eq.~\eqref{eqn:tLt_C2}, $\norm{\mathbf{l}_X-\mathbf{p}}^4\norm{\mathbf{l}_{Y}}^4\leq\norm{\mathbf{l}_X}^4\norm{\mathbf{l}_{Y}}^4$, which indeed proves Eq.~\eqref{eqn:to_prove}. Furthermore, it is evident that the matrix in Eq.~\eqref{eqn:tLt_C2} converges to $\mathbf{D}_{rms}$ in Eq.~\eqref{eqn:Drms_mat} when the orthogonality, $\mathbf{l}_{Y} \perp \mathbf{l}_{X}$, is fulfilled. Since in practice orthogonality is approached, but never fulfilled exactly, the statement that the RMS approximation always underestimates SEFD is demonstrably correct.

\textcolor{black}{We note the fact that antenna length is a direction dependent quantity. Therefore orthogonality, $\mathbf{l}_{Y} \perp \mathbf{l}_{X}$, as discussed in this section must be evaluated by taking the inner product, $\mathbf{l}_{X}^H\mathbf{l}_{Y}$, in the every direction of arrival of interest. This is particularly applicable to a wide field-of-view antenna system. For example, a mechanically orthogonal cross dipoles only possess antenna length orthogonality in directions of arrival on the cardinal planes (as discussed in Paper~I). The RMS approximation is only exact in directions of arrivals for which antenna length orthogonality, $\mathbf{l}_{X}^H\mathbf{l}_{Y}=0$, is fulfilled. }

Following the proof, we can then quantify this difference by defining a relative percentage difference of the two results as:
\begin{eqnarray}
\Delta \text{SEFD}_I =\frac{\text{SEFD}_{I}- \text{SEFD}_{I}^{rms}}{\text{SEFD}_{I}} \times 100 \% \geq 0.
\label{eqn:SEFD_I_error}
\end{eqnarray}
We will use  Eq.~\eqref{eqn:SEFD_I_error}  in Sec.~\ref{sec:compare} to measure the difference between RMS approximation and true SEFD based on simulation and observation, the expected value of which are always positive (for observation data, the ``expected value'' refers to the ensemble mean).

\section{SEFD simulation procedure}
\label{sec:simulation}
\begin{table*}[t]
\centering
\caption{ Calculated parameters required to compute $\text{SEFD}_I$ for the array pointing at Hydra-A source (Az=81\degree, ZA=46\degree) on 2014-12-26 at 16:05:43  UTC. \label{tab:values}}
\renewcommand{\arraystretch}{1.5}
\begin{tabular}{l|ccc||ccc|}
\cline{2-7}
       & \multicolumn{3}{c||}{\textbf{X}}                    & \multicolumn{3}{c|}{\textbf{Y}}         \\ \cline{2-7} 
       & \multicolumn{1}{c|}{110.08 MHz}  & \multicolumn{1}{c|}{154.88 MHz}  & \multicolumn{1}{c||}{249.6 MHz} & \multicolumn{1}{c|}{110.08 MHz}  & \multicolumn{1}{c|}{154.88 MHz}  & \multicolumn{1}{c|}{249.6 MHz} \\ \hline
\multicolumn{1}{|l||}{$T_{ant}~(\si{\kelvin})$}   & \multicolumn{1}{c|}{603} & \multicolumn{1}{c|}{227} & 66          & \multicolumn{1}{c|}{573} & \multicolumn{1}{c|}{222} & 63       \\ \hline
\multicolumn{1}{|l||}{$T_{\mathrm{rcv}}~(\si{\kelvin})$}   & \multicolumn{1}{c|}{180} & \multicolumn{1}{c|}{80} & 59          & \multicolumn{1}{c|}{126} & \multicolumn{1}{c|}{51} & 50       \\ \hline
\multicolumn{1}{|l||}{$\eta_{\mathrm{rad}}$}   & \multicolumn{1}{c|}{0.986} & \multicolumn{1}{c|}{0.972} & 0.989          & \multicolumn{1}{c|}{0.981} & \multicolumn{1}{c|}{0.980} & 0.951       \\ \hline
\multicolumn{1}{|l||}{$R~(\Omega)$}      & \multicolumn{1}{c|}{15}    & \multicolumn{1}{c|}{51}    &  {71}    & \multicolumn{1}{c|}{22}    & \multicolumn{1}{c|}{113}    &  {81}   \\ \hline
\multicolumn{1}{|l||}{$l_\theta~(\si{\metre})$} & \multicolumn{1}{c|}{$0.112 +j1.441$}    & \multicolumn{1}{c|}{$2.372 - j0.113$}    &   $-0.095 + j1.931$    & \multicolumn{1}{c|}{$0.022 + j0.237$}    & \multicolumn{1}{c|}{$0.441 - j0.106$}    & $0.008 + j0.740$  \\ \hline
\multicolumn{1}{|l||}{$l_\phi~(\si{\metre})$}  & \multicolumn{1}{c|}{$-0.008 - j0.299$}    & \multicolumn{1}{c|}{$-0.494 + j0.018$}    &  $0.080 - j0.339$       & \multicolumn{1}{c|}{$0.083 + j1.962$}    & \multicolumn{1}{c|}{$3.779 - j0.680$}    & $-0.880 + j2.045$   \\ \hline
\end{tabular}
\end{table*}

Following the previous section, the steps to computing the SEFD of a polarimetric phased array interferometer are as follows. The first step involves the computation of the array Jones matrix as shown in Eq.~\eqref{eqn:J_array}. The components of the matrix were constructed from embedded antenna realized lengths as shown in Eq.~\eqref{eqn:array_lX} and Eq.~\eqref{eqn:array_lY}, which in our case were obtained from electromagnetic simulation using Altair FEKO\footnote{https://www.altair.com/feko/}. Additional information regarding the simulation setup and results used in this paper were discussed in \citet{sokolowski_colegate_sutinjo_2017}. The embedded antenna realized lengths were found by converting $(E_{\theta}, E_{\phi})$ using \citep{Ung_9040892}
\begin{eqnarray}
l_{n\theta}(\theta,\phi) &=& -j \frac{2\lambda}{\eta_0 V_{t}}Z_{LNA}E_{\theta,n}(\theta,\phi), \nonumber\\ 
l_{n\phi}(\theta,\phi) &=&  -j \frac{2\lambda}{\eta_0 V_{t}}Z_{LNA}E_{\phi,n}(\theta,\phi),
\label{eqn:lr_calc}
\end{eqnarray}
where $\lambda$ is the wavelength, $Z_{LNA}$ is the input impedance of the low-noise amplifier (LNA), and $V_{t}$ is the port excitation voltage used during the simulation which generates the corresponding electric far-field components $E_{\theta,n}$ and $E_{\phi,n}$ of the $n^{th}$ embedded element as a function of direction in the sky. 

In the second step, we determine the system temperatures of the array, $T_{sysX}$ and $T_{sysY}$ using 
\begin{eqnarray}
T_{sys,p} = T_{ant,p} + T_{rcv,p} + T_0(1-\eta_{rad,p}),
\label{eqn:Tsys_formula}
\end{eqnarray}
where $T_{ant,p}$ is the antenna temperature due to sky for each polarization $p$ ($p$~=~``X'' or ``Y''); $T_{rcv,p}$ is the receiver noise temperature due to the LNA calculated using the methodology discussed in~\citet{Ung_9040892} based on measured LNA noise parameters~\citep{Ung2018}; $\eta_{rad,p}$ is the radiation efficiency of the array and $T_0 = 290$~\SI{}{\kelvin} is the reference temperature (this can be substituted by ambient temperature if known). Radiation efficiency, $\eta_{\mathrm{rad}}$, is the ratio of the total radiated power to the total injected power into the array calculated using methodologies presented in~\citet{5439763} and~\citet{8888935}.

Third, we compute the realized array noise resistances, $R_X$ and $R_Y$, using Eq.~\eqref{eqn:R_noise}. Fourth, we calculate $\mathbf{t}_R^T \mathbb{L}\mathbf{t}_R$ in Eq.~\eqref{eqn:var_I_simp_array}. Finally, we calculate the polarimetric and interferometric array SEFD using Eq.~\eqref{eqn:SEFD_I_res_array}. As an example, we provided several key parameters required to evaluate Eq.~\eqref{eqn:SEFD_I_res_array} at three selected frequency points for an MWA tile pointed at azimuth and zenith angles, Az $= 80.54^\circ$, ZA$= 46.15^\circ$ in Table \ref{tab:values}. A full list of these values (excluding the realized length) can be found in the dataset accompanying this paper. In the next section, we will describe the procedure used to obtain $\text{SEFD}_{I}$ from observational data.


\section{SEFD measurement procedure}
\label{sec:obs}

%

The data analysis procedure is based on a very similar analysis performed earlier in the Paper~I. The relative difference $\Delta \text{SEFD}_I^{obs}$ defined by Eq.~\eqref{eqn:SEFD_I_error} was measured using MWA observations tracking Hydra-A radio-galaxy recorded in a standard observing mode at the central frequency 154.88\,MHz. On 2014-12-26 between 16:05:42 and 05:13:42 UTC, the MWA recorded 23 112\,s observations centred on the Hydra-A radio-galaxy at a position (Az,ZA) = (80.7\degree, 45.8\degree) and (Az,ZA) = (290.7\degree, 32.4\degree) at the start and end of the observations respectively. 
For the purpose of a comparison of $\Delta \text{SEFD}_I$ between the data and simulations, 52\,s of the first 112\,s observation (obsID 1103645160 referred later as ObsA), which started at 2014-12-26 16:05:42 UTC with Hydra-A at (Az,ZA) = (80.7\degree, 45.8\degree), were analysed. This observation was selected at a pointing direction where the simulation predicted that $\Delta \text{SEFD}_I$ would be the most prominent due to the lowest elevation of the pointing direction at gridpoint 114 at the elevation of $\approx~$43.85\degree (ZA = 46.15\degree). We note that MWA observations significantly lower than 45$\degree$ are generally not routinely performed and were not considered for the presented analysis due systematic effects which could affect the data quality.

\subsection{Calibration and imaging}
\label{ref:calibration_and_imaging}

The MWA data were converted into CASA measurement sets \citep{casa} in 40\,kHz frequency (total of 768 channels) and 1\,s time resolution, and downloaded using the MWA All-Sky Virtual Observatory interface \citep{asvo}. In order to avoid aliasing effects at the edges of 1.28\,MHz coarse  channels (24 in total), 160\,kHz (4 fine channels) on each end of coarse channels were excluded, which reduced the observing bandwidth to $R_{bw}=$0.75 fraction of the full recorded band (30.72\,MHz). 

\begin{figure*}
\centering
  \includegraphics[width=\textwidth]{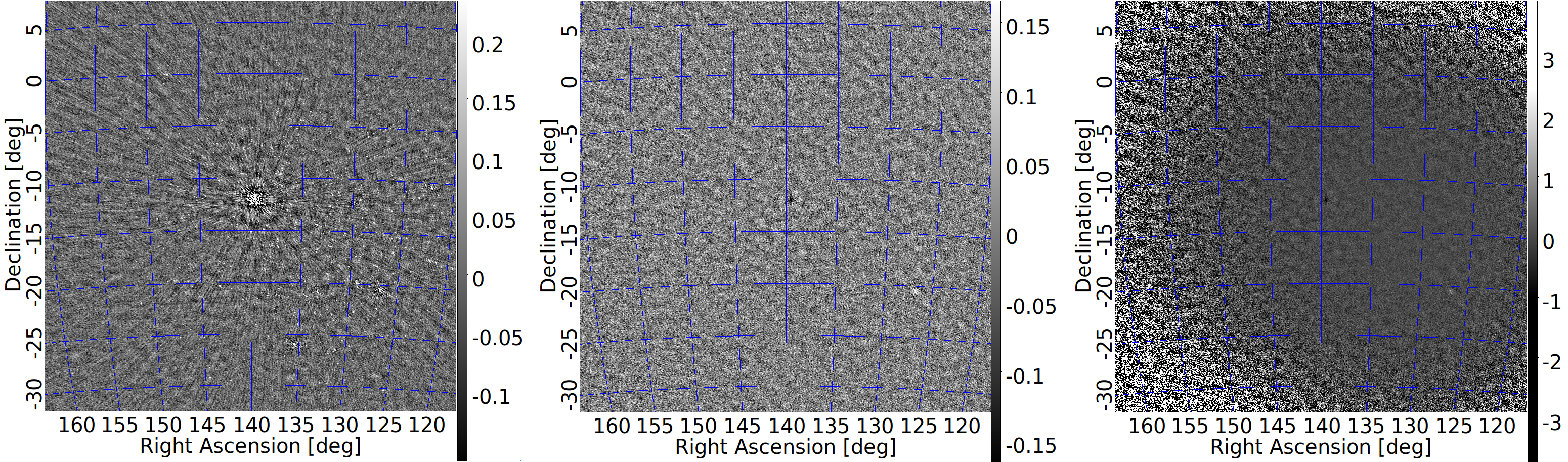}
\caption{Example 0.5\,s sky and difference images in X polarisation (scale of the colorbar is in Jy) used to calculate SEFD. Left : 0.5\,s sky image. Centre: difference of two consecutive 0.5\,s images. Right: difference of two consecutive 0.5\,s images corrected for the primary beam (divided by the primary beam in the X polarisation). In order to calculate noise in every direction in the sky within the field of view (``noise map''), standard deviation of the noise was calculated in small regions around each pixel in the difference image and divided by $\sqrt{2}$. These noise maps were later converted into SEFD using equation~\ref{eq_rms2sefd}. The same procedure was applied to Y and very similar to Stokes I polarisations (in this case the correction for the primary beam was more complex than the simple division by the corresponding primary beam).}
\label{fig:sky_images}
\end{figure*}

The same observation ObsA was used for calibration. It was sufficient to use a single-source sky model composed of Hydra-A alone, which was the dominant source in the observed field. However, due to the extended structure of Hydra-A, which is resolved by the longest baselines of the MWA Phase 1, it was necessary to use its model by \citet{2014MNRAS.440..327L} converted to the required multi-component format by \citet{gleamx}\footnote{https://github.com/nhurleywalker/GLEAM-X-pipeline}. The total flux density of Hydra-A in this model is within 7\% from the most recent and accurate flux measurements of Hydra-A by \citet{2017ApJS..230....7P} over the entire observing band (139.5 -- 169.0\,MHz). Selecting an appropriate model of Hydra-A was important to ensure correct flux scale of the calibrated visibilities, images and ultimately SEFD values.
The calibration was performed and applied using \textsc{CALIBRATE} and \textsc{APPLYSOLUTIONS} software, which are part of the MWA-reduce software package \citep{offringa-2016} routinely used for the MWA data reduction. The \textsc{calibrate} program uses the 2016 MWA beam model \citep{sokolowski_colegate_sutinjo_2017} to calculate Jones matrices and apparent flux of the sources (just one in this particular case) included in the sky model. Baselines shorter than 58 wavelengths were excluded from the calibration. The phases of the resulting calibration solutions were fitted with a linear function, while the amplitudes were fitted with a $5^{th}$ order polynomial; the resulting fitted calibration solutions were then applied to the un-calibrated visibilities using \textsc{APPLYSOLUTIONS} program.

Sky images with 4096$\times$4096 pixels of angular size $\approx$0.6\,arcmin (image size $\approx$\,42.2\degree$\times$42.2\degree), in 1\,s time resolution, were formed from all correlations products ($XX, YY, XY$ and $YX$)\footnote{We note that the correlation products are in fact $XX^*, YY^*, XY^*$ and $YX^*$, but in the context of this section the conjugate symbols ($^*$) were dropped for brevity.} using the \textsc{WSCLEAN}\footnote{\url{https://sourceforge.net/p/wsclean/wiki/Home/}} program \citep{OffMcK14}. Natural weighting (robust weighting parameter defined by \citet{Briggs_phd_thesis} set to $+2$) was used in order to preserve full sensitivity of the array. We note that natural weighting increases the confusion noise, but it was eliminated by measuring the noise from difference images (see Sec.~\ref{ref:noise2sefd}). The dirty maps were CLEANed with 100000 iterations and threshold of 1.2 standard deviations of the local noise. 
In order to reduce contribution from extended emission baselines shorter than 30 wavelengths were excluded, which led to a small reduction of number of baselines used in the imaging to 7466, which is $R_{bs} \approx$0.92 fraction of all the baselines (8128). Corrections for reduced number of baselines and observing bandwidth were taken into account in the later analysis by applying correction factors ($R_{bw}$ and $R_{bs}$) during the conversion from standard deviation of the noise to SEFD. 

The resulting $XX$ and $YY$ images (for example left image in Fig.~\ref{fig:sky_images}) were divided by corresponding images of the beam in X and Y polarizations generated with the 2016 beam model \citep{sokolowski_colegate_sutinjo_2017} at the same pointing direction ( (Az,ZA) = (80.5\degree , 46.2\degree) at the MWA gridpoint 114\footnote{MWA gridpoints are pointing directions where delays applied in the analogue beamformers are exact.}).
All the resulting images ($XX, YY, XY$ and $YX$) were converted to Stokes images ($I, Q, U$ and $V$) also using the 2016 beam model. The final products resulting from the above procedure were three sets ($XX, YY$ and Stokes $I$) of $n_t=53$ primary beam corrected images corresponding to  53\,s of the analysed MWA data.

\subsection{Measuring the SEFD from the noise in the sky images}
\label{ref:noise2sefd}

Using the three series (corresponding to $XX$, $YY$ and Stokes $I$) of $n_t=53$ images, difference images (between the subsequent $i$-th and $i-1$ image) were calculated, resulting in $n_d=52$ difference images in each of $XX$, $YY$ and Stokes $I$ polarizations. Difference imaging effectively removes confusion noise resulting in uniform images of thermal noise (entirely due to the system temperature $T_{sys}$). The difference images were visually inspected and verified to have a uniform noise-like structure (examples in the middle and right images in Fig.~\ref{fig:sky_images}). Therefore, the resulting standard deviation is purely due to the instrumental and sky noise ($T_{sys}$). 

The standard deviation of the noise calculated in a small region (here a circular region of radius $R_n$=10 pixels corresponding to $\approx$~5 synthesized beams) around pixels in the resulting difference images is distributed around a mean of zero and is not ``contamined'' by the variations of the flux density within the circular regions due to astronomical sources contained inside these regions. The interquartile range divided by 1.35 was used as a robust estimator of standard deviation, which is more robust against outlier data points due to radio-frequency interference (RFI), or residuals of astronomical sources in the difference images. Once this calculation was performed around all the pixels in a difference image, an image of the noise over the field of view (FoV) was created. These images of the noise are later referred to as noise maps or just N (note that N stands for a 2D image and not a number). This procedure was applied to $n_d$ all-sky difference images resulting in $n_d$ noise maps ($N_{I}$, $N_{XX}$ and $N_{YY}$) for each of the polarizations (Stokes $I$, $XX$ and $YY$ respectively). 
The corresponding SEFD images (2D maps of SEFD) of the entire FoV were calculated from these noise maps (referred in general as $N$) according to the equation:
\begin{equation}
\text{SEFD}= N \sqrt{ \Delta \nu \Delta t N_b},
\label{eq_rms2sefd}
\end{equation}
where $\Delta t=1$\,s is the integration time, $\Delta \nu = 30.72 \times R_{bw}$\,MHz is the observing bandwidth corrected for reduced number of channels ($R_{bw}$) used in the analysis and $N_b = 7466$ is the number of baselines used in the imaging.


Noise maps in the three polarizations, $N_{I}$, $N_{XX}$ and $N_{YY}$, were converted into corresponding $\text{SEFD}_I$, $\text{SEFD}_{XX}$ and $\text{SEFD}_{YY}$ images using Eq.~(\ref{eq_rms2sefd}), resulting in $n_d$ SEFD images in each polarization (I, XX and YY). Then, median SEFD images $\overline{\text{SEFD}_I}$, $\overline{\text{SEFD}_{XX}}$ and $\overline{\text{SEFD}_{YY}}$ were calculated out of individual $n_d$ SEFD images (a median image of $\text{SEFD}_{I}$ is shown in Figure~\ref{fig:obs_sefd_i}). These median images were used to calculate $\Delta \text{SEFD}_I^{obs}$ as:
\begin{eqnarray}
\Delta \text{SEFD}_I^{\text{obs}} =   \frac{( \overline{\text{SEFD}_{I}} - \frac{1}{2}\sqrt{\overline{\text{SEFD}_{XX}}^2 + \overline{\text{SEFD}_{YY}}^2} )}{\overline{\text{SEFD}_{I}}}.
\label{eqn:ratio}
\end{eqnarray} 
In the next section, we provide the results derived from simulated and observed data, and discuss their implications. 


\subsection{SEFD as a function of frequency}
\label{ref:sefd_vs_frequency}

\begin{figure}[b]
\centering
  \includegraphics[width=0.48\textwidth]{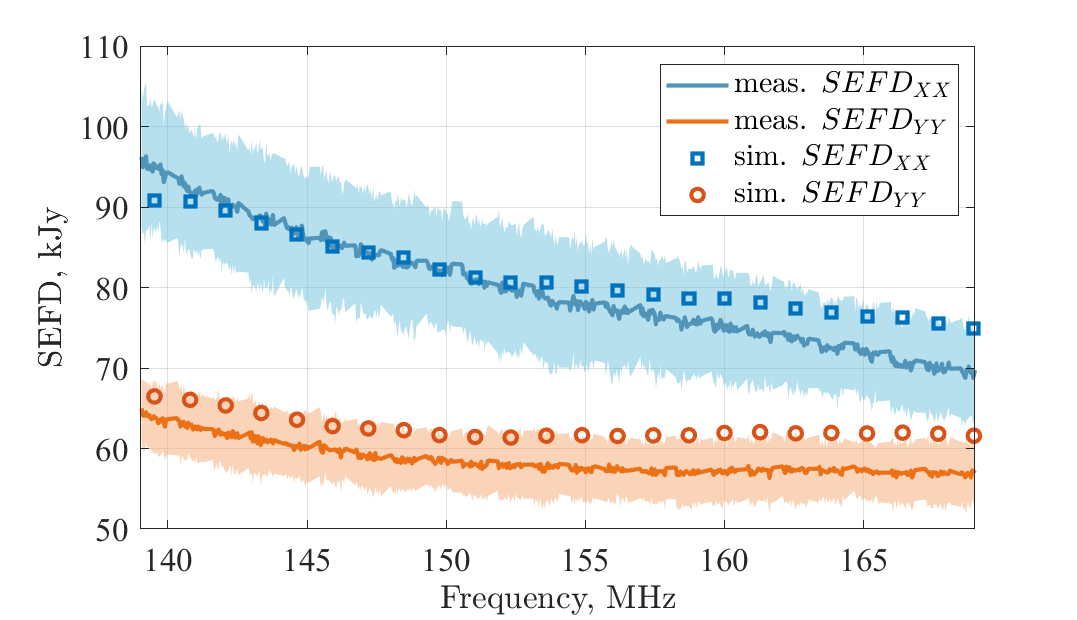}
\caption{SEFD \textcolor{black}{in the pointing direction of gridpoint 114 with realized area calculated at (Az, ZA) = (80\degree , 46\degree)} as a function of frequency. \textcolor{black}{Observed} SEFD calculated from calibrated visibilities compared with simulated SEFD.}
\label{fig:sefd_visib}
\end{figure}

The method described in Section~\ref{ref:calibration_and_imaging} allows us to determine spatial structure of SEFD by calculating SEFD values in each pixel of 4096$\times$4096 sky images. This method was successfully used in Paper I, where only single dipole in each MWA tile was enabled and the remaining 15 dipoles were terminated. However, it has never been applied to measure SEFD from MWA images resulting from observations in a standard mode with all dipoles in MWA tiles enabled. Therefore, before comparison with the simulations, an initial verification was performed by deriving SEFD values using well-tested method described in Section C of \citep{7293140}. This method uses standard deviation of calibrated visibilities to measure $\text{SEFD}_{XX}$ and $\text{SEFD}_{YY}$ (we highlight that only SEFD of instrumental polarizations can be measured using this method but not $\text{SEFD}_I$)  as a function of frequency in the direction of Hydra-A source (no spatial information). These measurements were used as a reference to confirm that the image-based SEFD values were correct. Furthermore, this method allowed us to compare frequency dependence of $\text{SEFD}_{XX}$ and $\text{SEFD}_{YY}$ against the simulations. Fig. \ref{fig:sefd_visib} shows the comparison between the observed and simulated results. \textcolor{black}{
At the location in the sky corresponding to (Az, ZA) = (80\degree, 46\degree), the realized area of X-polarized dipole is significantly smaller than Y-polarized dipole and therefore, $\text{SEFD}_{XX}$ is higher than $\text{SEFD}_{YY}$.}

At 154.88~MHz, observed $\text{SEFD}_{XX}=78\pm7.8$~kJy and $\text{SEFD}_{YY}=57.6\pm3.6$~kJy, which is in very good agreement with the values obtained from the imaging method. Overall, there is excellent agreement between simulated and observed $\text{SEFD}_{XX/YY}$ over a wide range of frequencies. 

\section{Results and discussion}
\label{sec:compare}
\subsection{Comparison between the data and simulation}

\begin{figure}[h!]
\centering
    \begin{subfigure}{.45\textwidth}
      \includegraphics[width=\linewidth]{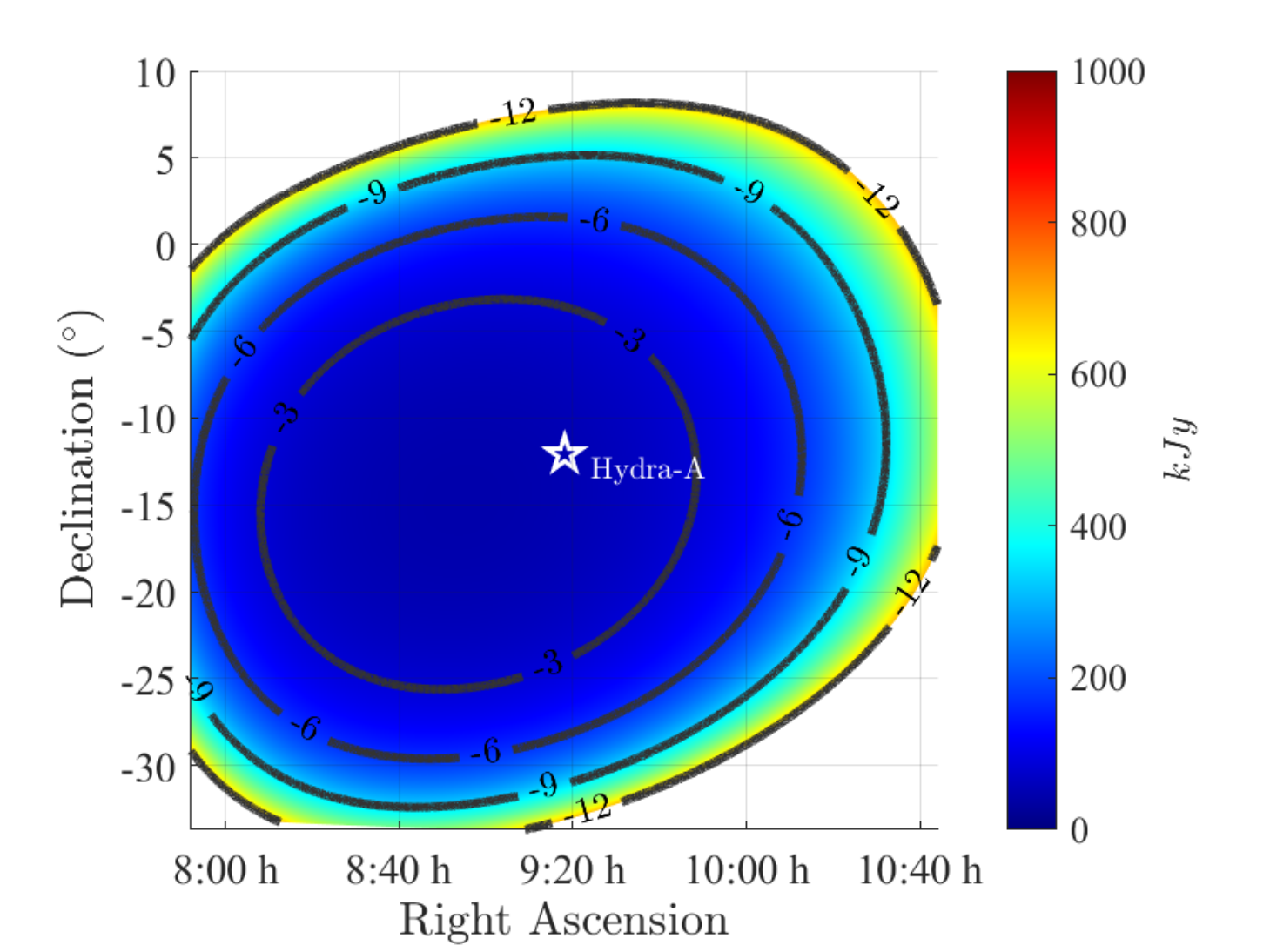}
      \caption{Simulated $\text{SEFD}_{I}$}
      \label{fig:sim_sefd_i}
      \end{subfigure}
        \begin{subfigure}{.45\textwidth}
      \includegraphics[width=\linewidth]{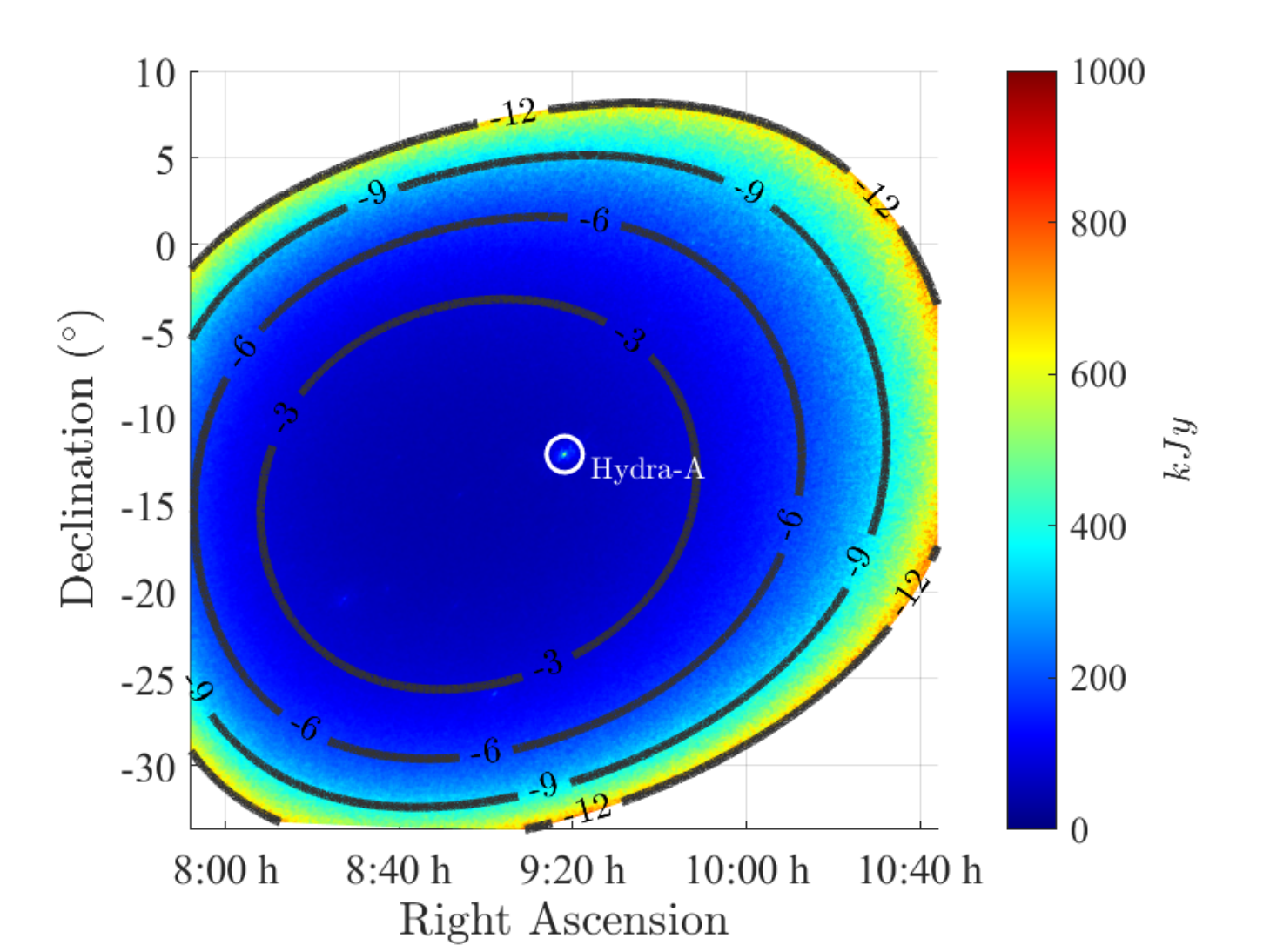}
      \caption{Observed $\text{SEFD}_{I}$}
      \label{fig:obs_sefd_i}
      \end{subfigure}
      \begin{subfigure}{.45\textwidth}
      \includegraphics[width=\linewidth]{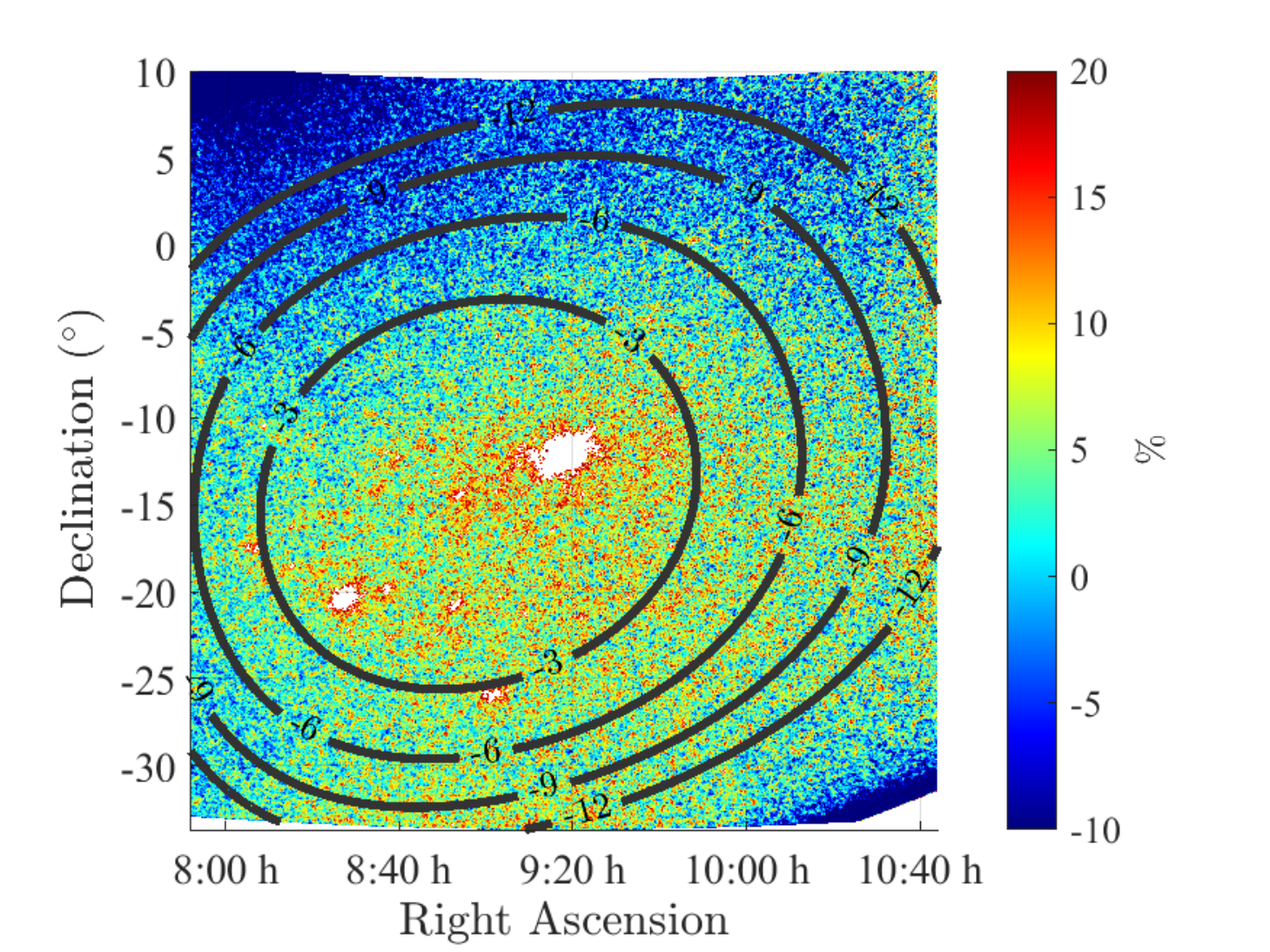}
      \caption{$M_{\text{SEFD}_{I}}$, \%, Eq.~(\ref{eqn:delta_s_m})}
     \label{fig:delta_sefd_i}
      \end{subfigure}
      \caption{(a) Simulated $\text{SEFD}_{I}$, (b) observed $\text{SEFD}_{I}$ and (c)~percentage difference between the simulated and observed $\text{SEFD}_{I}$ at 154.88~MHz calculated by Eq.~(\ref{eqn:delta_s_m}).} 
      \label{fig:sefd_sim_meas_comparI}
\end{figure}

In order to ensure that we are able to draw accurate conclusions regarding the difference in SEFD values calculated using the RMS approximation in Eq.~\eqref{eqn:SEFD_rms} and the true SEFD using Eq.~\eqref{eqn:SEFD_I_res_array}, we compared the simulated SEFD with the measurements. The SEFD of the $X,Y$ and $I$ using difference imaging method probes the SEFD over multiple locations in the sky for a given frequency (154.88 MHz). Similarly, the simulated $\text{SEFD}_{XX/YY}$ images were generated using $A/T$ formulation while the simulated $\text{SEFD}_{I}$ image was generated using Eq.~\eqref{eqn:SEFD_I_res_array}. 
Fig. \ref{fig:sefd_sim_meas_comparI} shows the SEFD obtained from simulation (Fig.~\ref{fig:sim_sefd_i}), Stokes I image (Fig.~\ref{fig:obs_sefd_i}) and the difference in percentage between the two  (Fig.~\ref{fig:delta_sefd_i}). For reference, contours of the normalized beam pattern for each polarization are displayed in $3$~dB increments down to $-12$~dB level. 
The contours of the beam pattern shown in the $\text{SEFD}_I$  are obtained by normalizing the reciprocal of Eq.~\eqref{eqn:SEFD_I_res_array}. Similar comparison was done for the X and Y polarization, but for brevity, the results are included in Appendix \ref{apendx:XXYY_comp}.

The relative difference in \%, named $M_{SEFD}$, between the simulated and observed SEFD in all images ($XX$, $YY$ and $I$) was computed as:
\begin{eqnarray}
M_{\text{SEFD}}= \frac{\text{SEFD}^{obs} - \text{SEFD}^{sim}}{\text{SEFD}^{sim}} \times 100 \%.
\label{eqn:delta_s_m}
\end{eqnarray} 
We analysed the errors using histograms and the observed $\text{SEFD}_{XX}$ and $\text{SEFD}_I$ are on average higher by 9\% and 4\% respectively, while the observed $\text{SEFD}_{YY}$ is lower by 4\% compared to simulated values for all pixels within the $-12$~dB beam width. Overall, the agreement between simulated and observed values is within the $\pm10\%$ range which indicates excellent correspondence of simulated and observed SEFD data.  

\begin{figure}[htb!]
\centering
          \begin{subfigure}{.45\textwidth}
      \includegraphics[width=\linewidth]{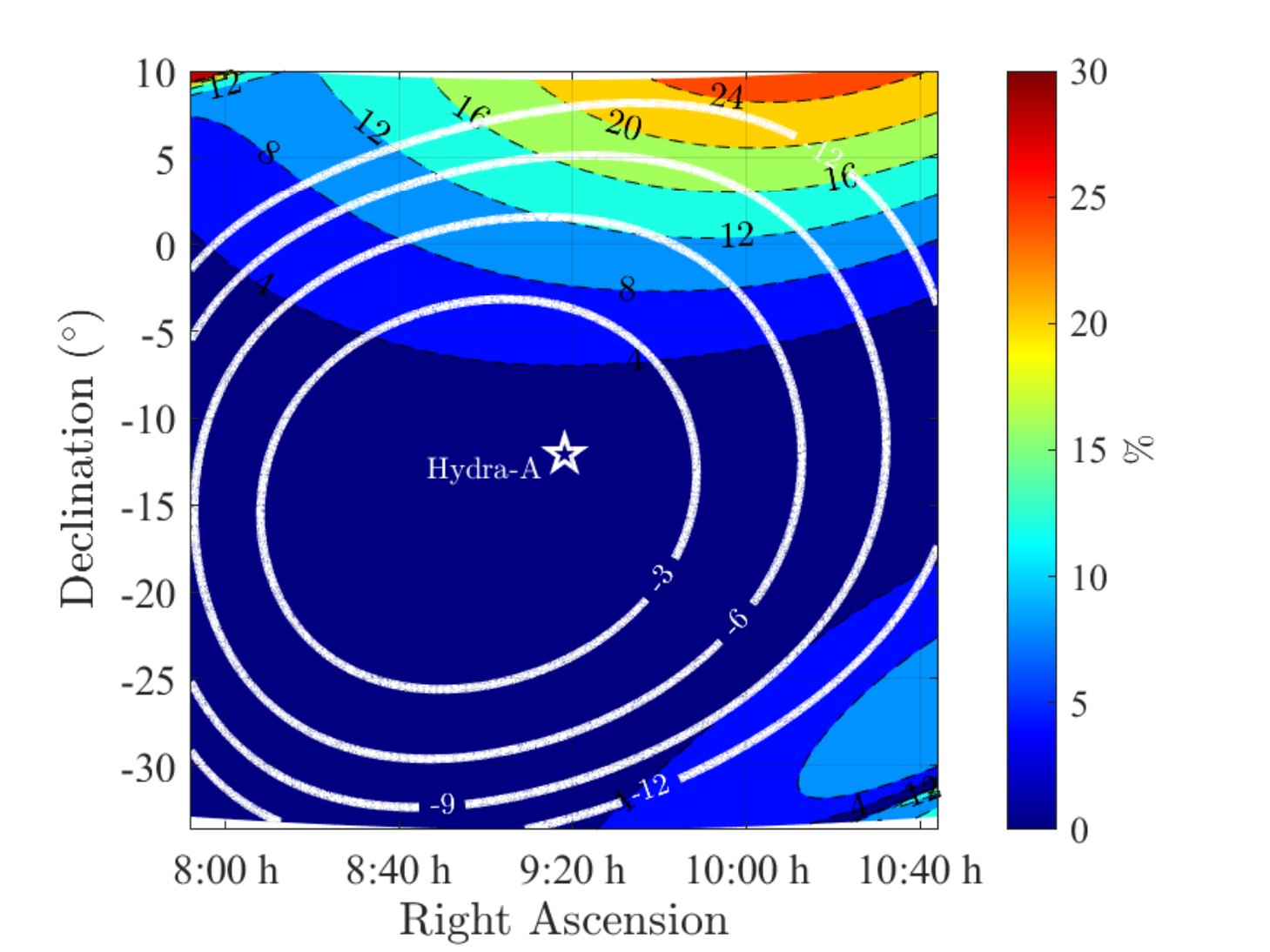}
      \caption{$\Delta \text{SEFD}_{I}^{sim}$, Eq.~(\ref{eqn:SEFD_I_error})}
      \label{fig:relative_ratio_simul}
      \end{subfigure}
            \begin{subfigure}{.45\textwidth}
      \includegraphics[width=\linewidth]{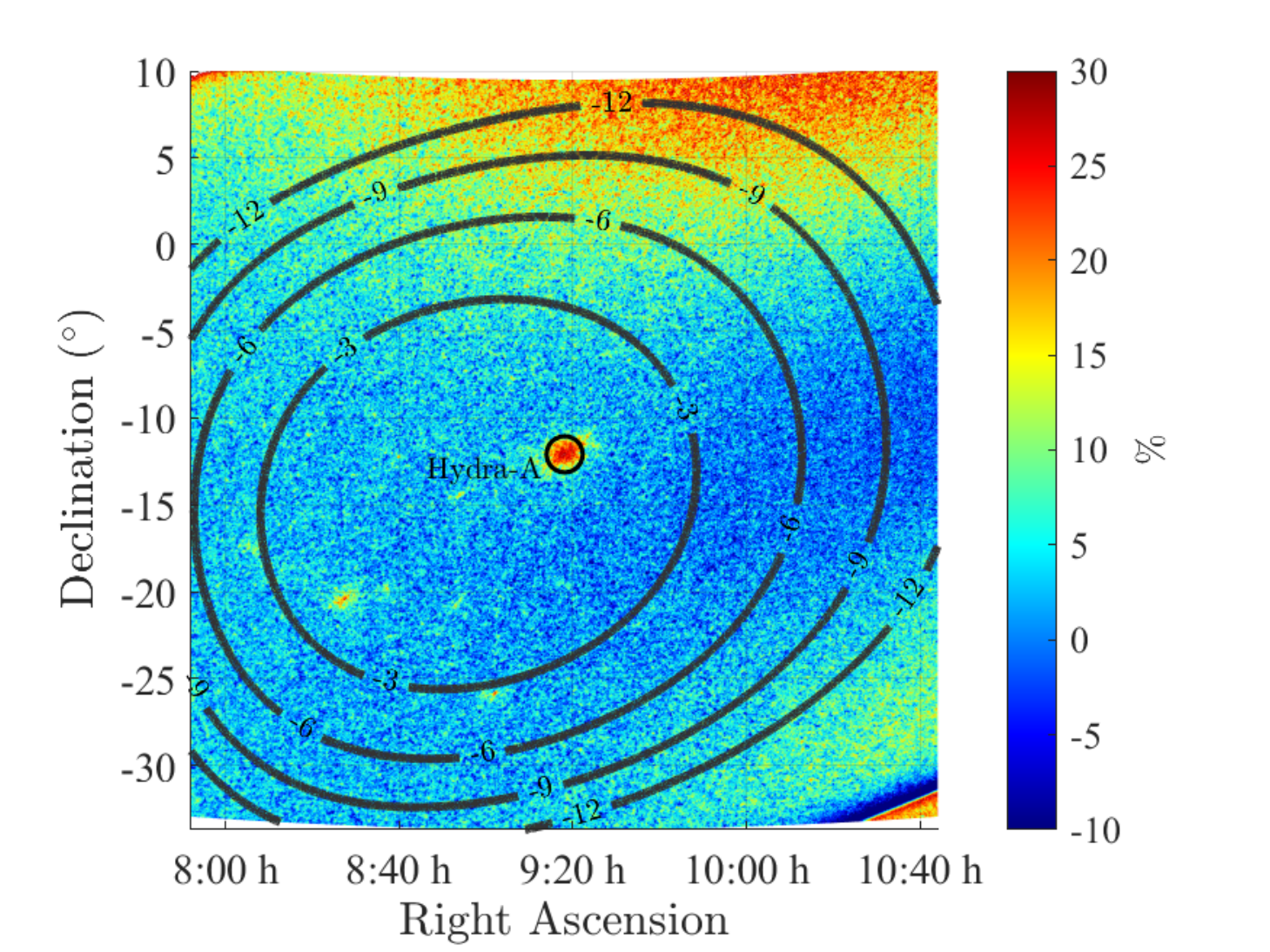}
      \caption{$\Delta \text{SEFD}_{I}^{obs}$, Eq.~(\ref{eqn:SEFD_I_error})}
      \label{fig:relative_ratio_data}
      \end{subfigure}
    \begin{subfigure}{.45\textwidth}
      \includegraphics[width=\linewidth]{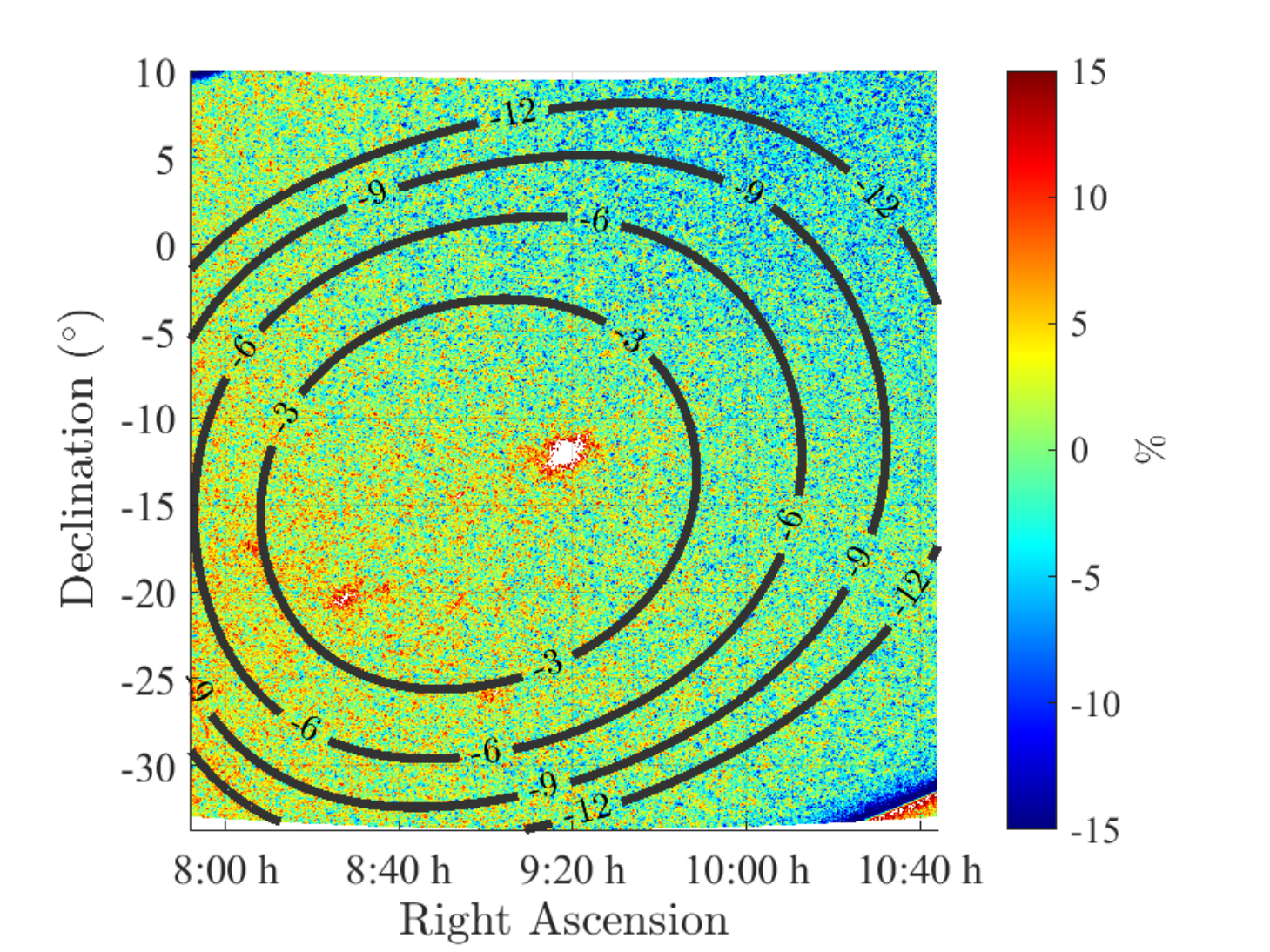}
      \caption{$\Delta_{sim/obs}$, Eq.~(\ref{eqn:delta_delta})}
      \label{fig:SEFDI_diff}
      \end{subfigure}
   \caption{Relative difference between the (a) simulated and (b) observed $\text{SEFD}_I$ and (c) the difference of $\Delta \text{SEFD}_I^{obs}$ and $\Delta \text{SEFD}_I^{sim}$.} 
      \label{fig:main_fig}
\end{figure}

Following the verification of simulated SEFD with observed data, we can proceed to calculate the error between the RMS and proposed method using Eq.~\eqref{eqn:SEFD_I_error}. Fig.~\ref{fig:relative_ratio_simul} and Fig.~\ref{fig:relative_ratio_data} show this error for an MWA tile for both simulated and observed data, respectively. We can see that the difference between $\text{SEFD}_I^{rms}$ and $\text{SEFD}_I$ is not a uniform offset in the image, but has a noticeable structure. For the observation within the $-3$~dB beam width, the error of SEFD prediction using the RMS approximation is only 7\%. However, it increases as we move further away from the beam center and reaches 23\% in the $-12$~dB beam width. We note that for error in the simulated values, $\Delta \text{SEFD}_{I}^{sim}$, is always positive, meaning the RMS method always underestimates the SEFD as expected.

Fig.~\ref{fig:SEFDI_diff} shows the subtraction of $\Delta \text{SEFD}_I^{obs}$ and $\Delta \text{SEFD}_I^{sim}$ calculated as 
\begin{eqnarray}
\Delta_{sim/obs} = \Delta \text{SEFD}_{I}^{obs} - \Delta \text{SEFD}_{I}^{sim}.
\label{eqn:delta_delta}
\end{eqnarray}
The resulting mean difference for all pixels within the $-12$~dB beam width is zero, which allows us to form the following conclusion: (i) the negative values seen in $\Delta \text{SEFD}_I^{obs}$ are due to noise and (ii) we can fully predict the amount of underestimation (\% errors) in $\text{SEFD}_I$ produced by the RMS method. This indicates that Eq.~\eqref{eqn:SEFD_I_res_array} more accurately predicts the array's $\text{SEFD}_{I}$.

\subsection{Computation of $\text{SEFD}_I$ for diagonal plane and low elevation pointing angle}
\begin{figure}[h!]
\begin{center}
\noindent
  \includegraphics[width=0.5\textwidth]{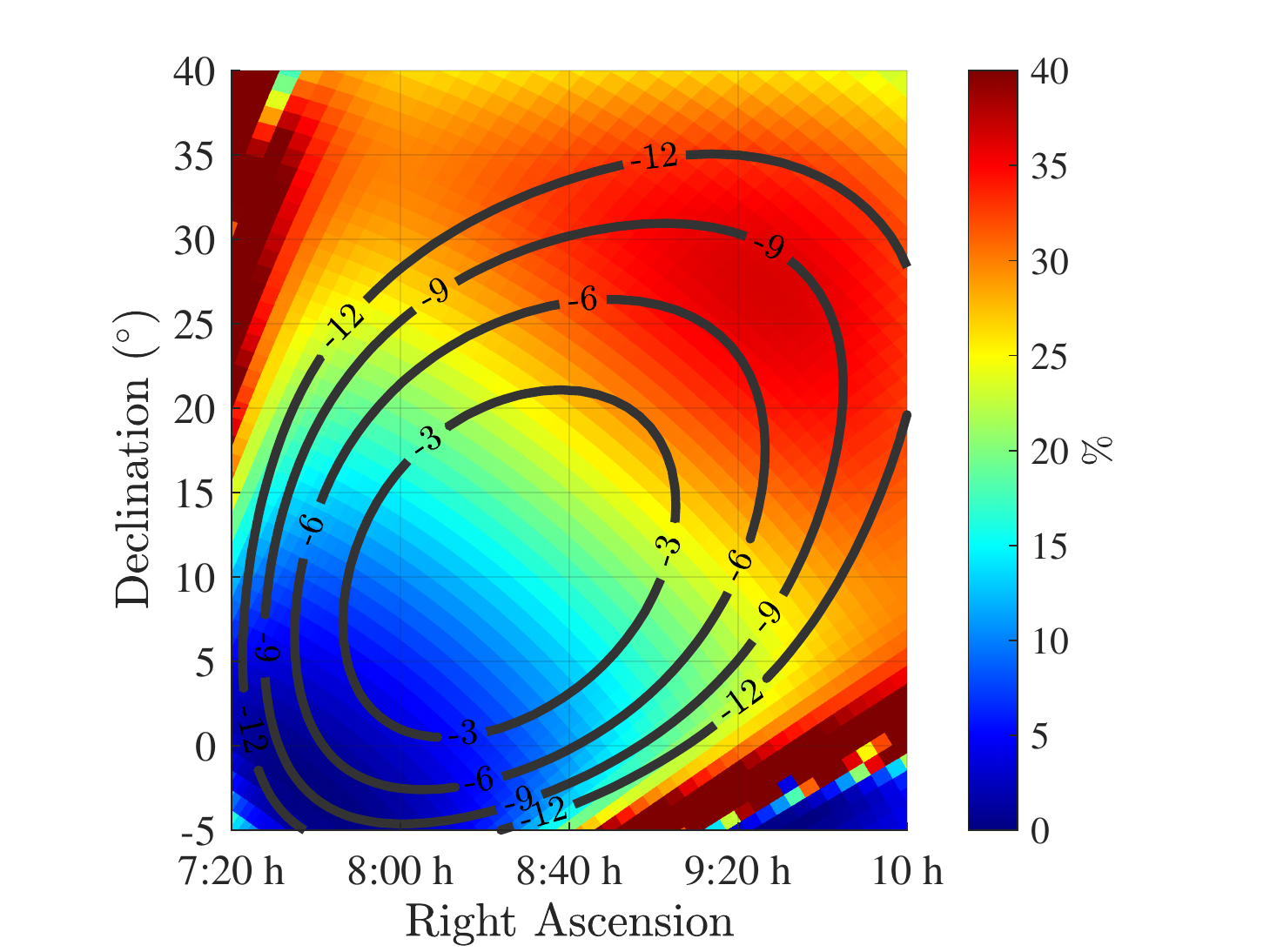}
\caption{$\Delta \text{SEFD}_I^{sim}$ computed for 199.68~MHz at (Az, ZA) = (45\degree, 56.96\degree), showing the impact of the error resulting from RMS expression of $\text{SEFD}_I$.}
\label{fig:worst_case}
\end{center}
\end{figure}
While there was a noticeable structure in the result shown in Fig. \ref{fig:main_fig}, we note that for that observation time, frequency and pointing angle, the difference in SEFD formulation would probably be hard to notice within the 3-dB beam width and indeed, in some cases, only images in the target field down to $-3$~dB are created. Nevertheless, these results show that the improved expression of $\text{SEFD}_I$ provides highly accurate SEFD values, and thus, can be used to precisely calculate the sensitivity of a radio telescope for any pointing angle within the operating frequency range using simulations.

To that end, it is reasonable to continue our analysis based only on simulated results. We completed an iterative search aimed to find a case where $\Delta \text{SEFD}_I$ affects the data within $-3$~dB beam width. Fig.~\ref{fig:worst_case} shows one outcome of this iterative search. It was found that for $\text{Az} = 45\degree$, $ZA=56.96\degree$  ($33.04 \degree $ elevation angle) at 199.68~MHz, the RMS SEFD approximation results in 29\% error within the $-3$~dB beam width and increases up to 36\% within the $-12$~dB beam width. This example further demonstrates the inaccuracies of the RMS approximation, which always predicts lower SEFD than the actual value.

\section{Conclusion}  \label{sec:concl}
This work, which is an extension of Paper~I, provides a generalized expression of $\text{SEFD}_I$ for interferometric phased array polarimeters. As in Paper~I, this expression was derived by performing statistical analysis on the standard deviation of the flux density estimate. Our current work further clarifies the fact that the $\text{SEFD}_I$ expression does not depend on assumptions made regarding the background polarization nor orthogonality in the antenna elements. The key array parameters were obtained from the full-wave electromagnetic simulation of the phased array of interest, which is an MWA tile in this example, and used to compute $\text{SEFD}_I$ given by Eq.~\eqref{eqn:SEFD_I_res_array}. The mean percentage difference ($M_{\text{SEFD}_{I}}$) between simulated and observed $\text{SEFD}_I$ is $4\%$, which indicates excellent agreement. Furthermore, we provided proof that the RMS approximation always underestimates the $\text{SEFD}_I$. This proof also demonstrates that the equality between the RMS approximation and $\text{SEFD}_I$ is reached when the row vectors in the Jones matrix are orthogonal, which is a condition that can be approached but never fulfilled in practice. 

After verifying the accuracy of our simulation with observed data, we proceeded to compare the relative percentage difference in $\text{SEFD}_I$, as defined in Eq.~(\ref{eqn:SEFD_I_error}), calculated by the often-used RMS approximation and Eq.~(\ref{eqn:SEFD_I_res_array}). A large percentage difference here would indicate that the RMS approximation underestimates the $\text{SEFD}_I$. Indeed, we see that for our chosen observation, the maximum simulated error $\Delta \text{SEFD}_{I}^{sim}=7\%$ within the $-3$~dB beam width and increases to $23\%$ within the $-12$ dB beam width. This result was also verified against the observed data ($\Delta \text{SEFD}_{I}^{obs}$) by taking absolute difference between the two images (Fig.~\ref{fig:relative_ratio_simul} and Fig.~\ref{fig:relative_ratio_data}) defined in Eq.~(\ref{eqn:delta_delta}). The resulting difference between simulated and observed $\Delta \text{SEFD}_I$ has a zero mean value, which indicates a remarkable agreement between the results. 

As we had thoroughly validated our simulation against observed data, we then performed an iterative search for pointing angles and frequencies that would yield a higher difference between the RMS approximation and Eq.~\eqref{eqn:SEFD_I_res_array}. We found that for beam pointing (Az, ZA) = (45\degree, 56.96\degree) at 199.68~MHz the RMS approximation produces an error of $29\%$ in $\Delta \text{SEFD}_{I}^{sim}$ within the $-3$~dB beam width, which increases to 36\% within the $-12$~dB beam width. This outcome is in agreement with the prediction made in Paper~I whereby the difference increases at $\text{Az} = 45\degree$ at low elevation angles. 

We conclude that the derived $\text{SEFD}_I$ expression improves the fundamental understanding of instrument performance and can be used to accurately calculate sensitivity not only at the principal planes, but also  at the diagonal planes (Az$=45\degree$) and low elevation angles. This enables us to more confidently predict sensitivity for detection of pulsars, fast radio bursts (FRBs) or Epoch of Reionisation (EoR) signal. This is particularly important for the cases where the target sources can only be observed at very low elevations (e.g., in effect of response to alerts, about FRBs or any other transient sources). 

\section*{Acknowledgement}
\label{sec:Ack}
This scientific work makes use of the Murchison Radio-astronomy Observatory (MRO), operated by CSIRO. We acknowledge the Wajarri Yamatji people as the traditional owners of the Observatory site. Support for the operation of the MWA is provided by the Australian Government (NCRIS), under a contract to Curtin University administered by Astronomy Australia Limited. This work was further supported by resources provided by the Pawsey Supercomputing Centre with funding from the Australian Government and the Government of Western Australia. The authors thank A/Prof. R. B. Wayth and Prof. D. Davidson for discussions on this topic and for reviewing the draft manuscript. The authors thank Dr. Natasha Hurley-Walker for providing the model of the Hydra-A calibrator source.

\bibliographystyle{aa}
\bibliography{Sens_array_17Mar2021.bib}

\begin{appendix}
\section{Statistical analysis of $\text{Var}(\tilde{I})$}
\label{apendx:stat}
The statistical reasoning and the vanishing covariance are similar to the appendix in~Paper~I. In this paper, we were able to reduce the assumptions only to the most fundamental ones, namely zero mean Gaussian noise with independent and identically distributed real and imaginary parts (in phasor domain) and zero mutual coherence of system noise between two arrays forming a baseline. We do not assume the polarization of the sky, the orthogonality of the elements in the array, or the lack of correlation of noise complex noise sources.  

The right hand size of Eq.~\eqref{eqn:Isimp} has the form 
\begin{eqnarray}
W = aX_1X_2^*-zX_1Y_2^* -z^*Y_1X_2^*+bY_1Y_2^*,
\label{eqn:sum}
\end{eqnarray}
where $X_1,X_2, Y_1, Y_2$ are complex random variables representing the voltages seen at the outputs of the summer; $a=|l_{Y\phi}|^2+|l_{Y\theta}|^2, b=|l_{X\phi}|^2+|l_{X\theta}|^2$ are real constants and $z=l_{X\phi}^*l_{Y\phi}+l_{X\theta}^*l_{Y\theta}$ is a complex constant. The variance is
\begin{eqnarray}
\text{Var}(W) &=& \text{Var}(aX_1X_2^*-zX_1Y_2^* -z^*Y_1X_2^*+bY_1Y_2^*),
\nonumber \\
&=&|a|^2\text{Var}(X_1X_2^*)+|z|^2\text{Var}(X_1Y_2^*)\nonumber \\
&+&|z|^2\text{Var}(Y_1X_2^*)+|b|^2\text{Var}(Y_1Y_2^*)+2C,
\label{eqn:varW}
\end{eqnarray}
where $C$ is the covariance of cross terms. The conjugation signs ($\_^*$ ) have been included here for clarity in the $\text{Var}(X_1 X_2^*)$ term et cetera. We begin by considering this term 
\begin{eqnarray}
\text{Var}(X_1X_2^*)&=&\left<\left|X_1X_2^*-\left<X_1X_2^*\right>\right|^2\right> \\
&=&\left<\left(X_1X_2^*-\left<X_1X_2^*\right>\right)\left(X_1^*X_2-\left<X_1^*X_2\right>\right)\right> \nonumber \\
&=&\left<X_1X_2^*X_1^*X_2\right>-\left<X_1X_2^*\left<X_1^*X_2\right>\right> \nonumber\\
&&-\left<\left<X_1X_2^*\right>X_1^*X_2\right> +\left<\left<X_1X_2^*\right>\left<X_1^*X_2\right>\right>. \nonumber
\label{eqn:var1a}
\end{eqnarray}
We recognize that the last two terms of the last line cancel because
\begin{eqnarray}
\left<\left<X_1X_2^*\right>X_1^*X_2\right>&=&\left<X_1X_2^*\right>\left<X_1^*X_2\right> \nonumber \\ \left<\left<X_1X_2^*\right>\left<X_1^*X_2\right>\right> &=& \left<X_1X_2^*\right>\left<X_1^*X_2\right>.
\label{eqn:var1_cancel}
\end{eqnarray}
This leaves us with
\begin{eqnarray}
\text{Var}(X_1X_2^*)= \left<X_1X_2^*X_1^*X_2\right>-\left<X_1X_2^*\right>\left<X_1^*X_2\right>. 
\label{eqn:var1b}
\end{eqnarray}
For the first term in the right hand side of Eq.~\eqref{eqn:var1b} we apply the formula for zero-mean joint Gaussian random variable $Z_{1,2,3,4}$~\citep{Thompson2017_rx_sys, Baudin_App3}
\begin{eqnarray}
\left<Z_1Z_2Z_3Z_4\right>&=&\left<Z_1Z_2\right>\left<Z_3Z_4\right>
+\left<Z_1Z_3\right>\left<Z_2Z_4\right> \nonumber\\
&+&\left<Z_1Z_4\right>\left<Z_2Z_3\right>.
\label{eqn:Thompson_formula}
\end{eqnarray}
Therefore
\begin{eqnarray}
\left<X_1X_2^*X_1^*X_2\right>&=&\left<X_1X_2^*\right>\left<X_1^*X_2\right>+\left<\left|X_1\right|^2\right>\left<\left|X_2\right|^2\right> \nonumber \\
&+&\left<X_1X_2\right>\left<X_1^*X_2^*\right>.
\label{eqn:var1c}
\end{eqnarray}
The first term in the right hand side of Eq.~\eqref{eqn:var1c} cancels the last term in Eq.~\eqref{eqn:var1a}, leaving us with
\begin{eqnarray}
\text{Var}(X_1X_2^*)= \left<\left|X_1\right|^2\right>\left<\left|X_2\right|^2\right> 
+\left<X_1X_2\right>\left<X_1^*X_2^*\right>. 
\label{eqn:var1d}
\end{eqnarray}
The last term in the right hand side of Eq.~\eqref{eqn:var1d} contain $X_1X_2$ that are not conjugated with each other in the expectation operation $\left<.\right>$, where each $X_1$, $X_2$ represent complex noise. We now show that  We can write this as
\begin{eqnarray}
\left<XY\right>&=&\left<(\Re_X+j\Im_X)(\Re_Y+j\Im_Y)\right>\nonumber\\
&=&\left<\Re_X\Re_Y\right>-\left<\Im_X\Im_Y \right>+j\left<\Re_X\Im_Y\right>\nonumber\\
&+&j\left<\Re_Y\Im_X\right>.
\label{eqn:XY_iid_cplx}
\end{eqnarray}
For independent real and imaginary parts, the terms $\left<\Re_X\Im_Y\right>,\left<\Re_Y\Im_X\right>$ vanish. Furthermore, for identical  correlation in the real part and in the imaginary part, we have $\left<\Re_X\Re_Y\right>-\left<\Im_X\Im_Y \right>=0$. These are consistent with zero-mean Gaussian noise representing thermal noise. Under the foregoing conditions, $\left<X_1X_2\right>=\left<X_1^*X_2^*\right>=0$. This leaves us with the key result
\begin{eqnarray}
\text{Var}(X_1X_2^*)
&=&\left<|X_1|^2\right>\left<|X_2|^2\right>\nonumber \\
&=&(4k\Delta f)^2T_{sysX1}R_{X1}T_{sysX2}R_{X2}. 
\label{eqn:varsX1X2}
\end{eqnarray}
With the same reasoning, it can be shown that
\begin{eqnarray}
\text{Var}(X_1Y_2^*)&=&\left<|X_1|^2\right>\left<|Y_2|^2\right> \nonumber \\
&=&(4k\Delta f)^2T_{sysX1}R_{X1}T_{sysY2}R_{Y2}, \nonumber \\
\text{Var}(Y_1X_2^*)&=&\left<|Y_1|^2\right>\left<|X_2|^2\right> \nonumber \\
&=&(4k\Delta f)^2T_{sysY1}R_{Y1}T_{sysX2}R_{X2}, \nonumber \\
\text{Var}(Y_1Y_2^*)&=&\left<|Y_1|^2\right>\left<|Y_2|^2\right> \nonumber \\
&=&(4k\Delta f)^2T_{sysY1}R_{Y1}T_{sysY2}R_{Y2},
\label{eqn:vars_rest}
\end{eqnarray}
We note that only zero-mean joint Gaussian random variable and iid real and imaginary parts are needed to obtain the results thus far. Also, for antenna arrays of an identical design, it is reasonable to let $R_{X1}=R_{X2}=R_{X}$ and $T_{sysX1}=T_{sysX2}=T_{sysX}$, and similarly with Y. 

In Eq.~\eqref{eqn:vars_rest}, the terms $\text{Var}(Y_1X_2^*)$ and $\text{Var}(Y_1Y_2^*)$ contribute to the overall variance in Eq.~\eqref{eqn:varW}. The terms $\text{Var}(Y_1X_2^*)$ and $\text{Var}(Y_1Y_2^*)$ are as significant as $\text{Var}(X_1X_2^*)$ and $\text{Var}(Y_1Y_2^*)$, but the contributions of the former are scaled by $|z|^2$. The $|z|^2$ term, in turn, becomes increasingly appreciable in the diagonal scan plane of the phased array with decreasing elevation angles. Moreover, this $|z|^2$ term is neglected in the RMS approximation of SEFD, which contributes to the underestimate.

Next, we consider the $C$ term
\begin{eqnarray}
C &=& -az\text{Cov}(X_1X_2^*,X_1Y_2^*)-az^*\text{Cov}(X_1X_2^*,Y_1X_2^*)
\nonumber\\
&+&ab \text{Cov}(X_1X_2^*,Y_1Y_2^*) +|z|^2\text{Cov}(X_1Y_2^*,Y_1X_2^*)
\nonumber \\
&-&bz\text{Cov}(X_1Y_2^*,Y_1Y_2^*)-bz^*\text{Cov}(Y_1X_2^*,Y_1Y_2^*).
\label{eqn:C}
\end{eqnarray}
So far there is no difference with the corresponding expressions in~Paper~I. We consider the covariance terms next, starting with
\begin{eqnarray}
\text{Cov}(X_1X_2^*,X_1Y_2^*)=\left<X_1X_2^*X_1Y_2^*\right>-\left<X_1X_2^*\right>\left<X_1Y_2^*\right>.
\label{eqn:cov1a}
\end{eqnarray}
Again, applying Eq.~\eqref{eqn:Thompson_formula} to the first term in the right hand side of Eq.~\eqref{eqn:cov1a}, we get
\begin{eqnarray}
\left<X_1X_2^*X_1Y_2^*\right>&=&\left<X_1X_2^*\right>\left<X_1Y_2^*\right>
+\left<X_1X_1\right>\left<X_2^*Y_2^*\right>\nonumber\\
&+&\left<X_1Y_2^*\right>\left<X_2^*X_1\right>. 
\label{eqn:cov1b}
\end{eqnarray}
As a result,
\begin{eqnarray}
\text{Cov}(X_1X_2^*,X_1Y_2^*)&=&\left<X_1X_1\right>\left<X_2^*Y_2^*\right>+\left<X_1Y_2^*\right>\left<X_2^*X_1\right> \nonumber\\
&=&\left<X_1Y_2^*\right>\left<X_2^*X_1\right>.
\label{eqn:cov1c}
\end{eqnarray}
The last line is due to $\left<X_1X_1\right>=\left<X_2^*Y_2^*\right>=0$ as discussed in Eq.~\eqref{eqn:XY_iid_cplx}. Following the same reasoning
\begin{eqnarray}
\text{Cov}(X_1X_2^*,Y_1X_2^*) &=&\left<X_1X_2^*\right>\left<X_2^*Y_1\right>, \nonumber\\
\text{Cov}(X_1X_2^*,Y_1Y_2^*) &=&\left<X_1Y_2^*\right>\left<X_2^*Y_1\right>, \nonumber\\
\text{Cov}(X_1Y_2^*,Y_1X_2^*)&=&\left<X_1X_2^*\right>\left<Y_2^*Y_1\right>,\nonumber\\
\text{Cov}(X_1Y_2^*,Y_1Y_2^*)&=& \left<X_1Y_2^*\right>\left<Y_2^*Y_1\right>,\nonumber \\
\text{Cov}(Y_1X_2^*,Y_1Y_2^*)&=&\left<Y_1Y_2^*\right>\left<X_2^*Y_1\right>.
\label{eqn:cov_check2}
\end{eqnarray}
Putting these results together into Eq.~\eqref{eqn:C}
\begin{eqnarray}
C &=& -az\left<X_1Y_2^*\right>\left<X_2^*X_1\right>-az^*\left<X_1X_2^*\right>\left<X_2^*Y_1\right>
\nonumber\\
&+&ab\left<X_1Y_2^*\right>\left<X_2^*Y_1\right> +|z|^2\left<X_1X_2^*\right>\left<Y_2^*Y_1\right>
\nonumber \\
&-&bz\left<X_1Y_2^*\right>\left<Y_2^*Y_1\right>-bz^*\left<Y_1Y_2^*\right>\left<X_2^*Y_1\right>.
\label{eqn:C2}
\end{eqnarray}
This is the point where the second assumption is introduced. We note in Eq.~\eqref{eqn:C2} that we are left with factors with differing subscripts, $\left<\__1,\__2\right>$. These factors vanish assuming  independent zero mean noise in Arrays 1 and 2. This is reasonable for low-frequency phased array radio telescopes since the dominant Galactic noise decays rapidly for baselines of tens of wavelengths~\citep{7293140}.

\begin{figure}[htb]
\begin{center}
\noindent
  \includegraphics[width=0.4\textwidth]{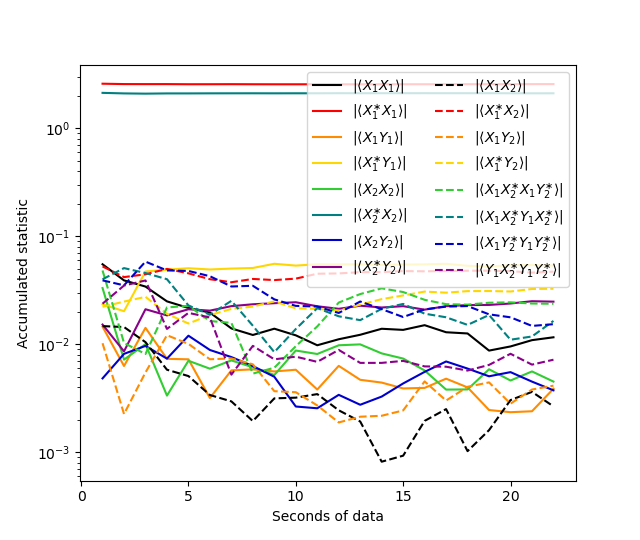}
\caption{The magnitudes of various cumulative statistics of data recorded with the MWA Voltage Capture System \cite{Tremblay2015} at two neighboring tiles. All except the auto-correlations $\left<|X|^2\right> = \left<X^*X\right>$ converge rapidly to zero.}
\label{fig:vcsstats}
\end{center}
\end{figure}

We validated our statistical reasoning by calculating various mean correlation products using the complex voltages measured at two neighboring MWA tiles (baseline $\approx 14\,$m), in both polarizations (i.e. $X_1$, $Y_1$, $X_2$, $Y_2$). The complex voltages were collected using the Voltage Capture System \citep{Tremblay2015}, at a frequency/time resolution of $10\,$kHz/$100\,\mu$s. The tile beam was pointed at Az/El of 51.3/40.6 deg. Fig.~\ref{fig:vcsstats} shows the cumulative statistics for one (arbitrary) channel, written out at a 1-second cadence. As predicted by the above formalism, the only significantly non-zero products are the auto-correlations; all other combinations converge rapidly to zero. In the final analysis $C=0$ in Eq.~\eqref{eqn:C} is well justified, which is the same conclusion as the appendix in~Paper~I. However, the reasoning here only involves the most fundamental assumptions as reviewed in this appendix.

\section{Linking $R$ to active reflection coefficient and $\tau$}
\label{apendx:links}
The quantity $R_{X}$ and $R_{Y}$ are the realized noise resistance of the array such that the noise voltage per $\sqrt{\Delta f}$ across the input impedance of the LNA can be calculated using $v_{n,X} = \sqrt{4kTR_{X}}$ and thus, the total power delivered per unit bandwidth is given by

\begin{eqnarray}
P_{del,\mathrm{ext}} = \Re\left\{\frac{|v_{n,X}|^2}{Z^*_{LNA}}\right\}.
\end{eqnarray}

The key point is that all the complex signal paths taken by the noise signal to reach the summing junction that exists due to mutual coupling are completely captured in $R_X$. Additionally, we can show that $R_X$ can be calculated using already established quantities such as $\tau_X$ and the active impedance $Z'_{actv,X}$ calculated from the active reflection coefficient.

The quantity $\tau$ derived in \cite{Ung_MPhil2020} is a ratio of the power delivered to the LNA input impedance due to the homogeneous sky to the available power ($kT$). Therefore, we can easily convert $\tau_X$ to $R_{X}$ and vice versa using
\begin{eqnarray}
\label{eqn:Rx_tau}
 R_{X}=\frac{\tau_X}{4\Re\left\{\frac{1}{Z^*_{LNA}}\right\}}.
\end{eqnarray}

Alternatively, the active reflection coefficient of an antenna element in an array as defined in \cite{Belo_7079488} can be used to calculate $R_{X}$. The active reflection coefficient accounts for all the coupling signal paths and combines it into a single equivalent signal path.   

Hence, the power for each equivalent branch can be calculated using
\begin{eqnarray}
P_{del,i} &=& \frac{4kT}{N}\left|\frac{Z_{LNA}\sqrt{\Re\{Z'_{actv,i}\}}}{Z_{LNA} + Z'_{actv,i}}\right|^2\Re\left\{\frac{1}{Z^*_{LNA}} \right\},
\end{eqnarray}
thus,
\begin{eqnarray}
\label{eqn:Rx_Zactv}
R_{X} &=& \frac{1}{N} \sum^{N}_{i=1} \left|\frac{Z_{LNA}}{Z_{LNA} + Z'_{actv,i}}\right|^2 \Re\{Z'_{actv,i}\}.
\end{eqnarray}
where $N$ is the number of antenna elements, and $Z'_{actv,i}$ is the active impedance for a given polarization obtained from the active reflection coefficient of the $i^{th}$ element. 

\begin{figure}[h!]
\begin{center}
\noindent
  \includegraphics[width=0.5\textwidth]{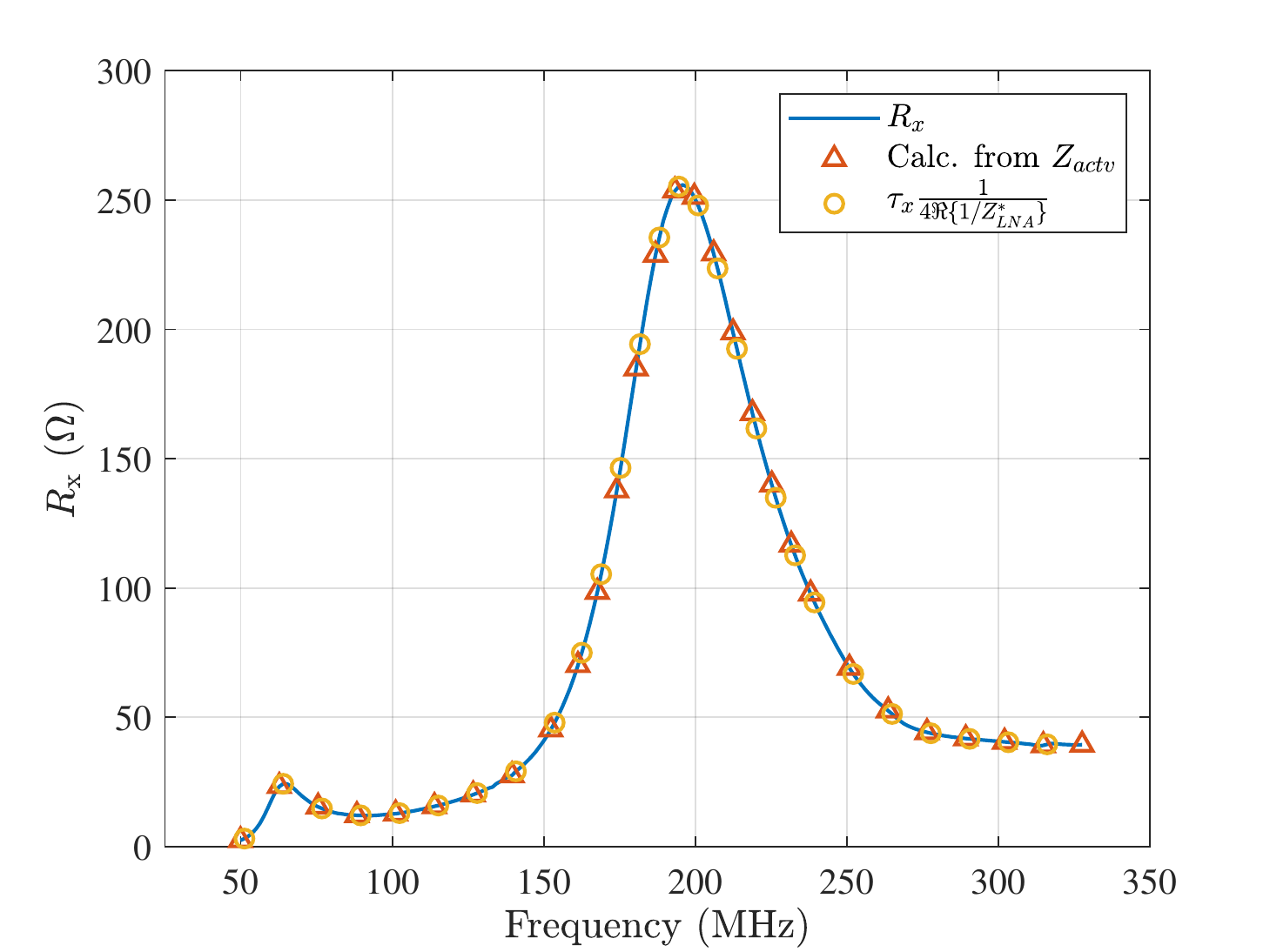}
\caption{Comparison of $R_{\mathrm{x}}$ calculated using three methods. The solid curve was obtained using equation~\eqref{eqn:R_noise}, while the data points represented by the circle and triangle markers are obtained using \eqref{eqn:Rx_tau} and \eqref{eqn:Rx_Zactv}, respectively.}
\label{fig:Rx_comp}
\end{center}
\end{figure}

Fig.~\ref{fig:Rx_comp} shows $R_{X}$ as a function of frequency calculated using three methods. Equations~\eqref{eqn:Rx_tau} and \eqref{eqn:Rx_Zactv} yield the same quantity as equation~\eqref{eqn:R_noise}. 

\section{Relation between RMS approximation and A/T}
\label{apendx:RMS2L}
As demonstrated by \citet{Ung_MPhil2020}, $\text{SEFD}_{XX/YY}$ can be calculated using realized area, $A_r$, thus, 
\begin{eqnarray}
\text{SEFD}_I^{rms}&=& \frac{1}{2}\sqrt{\text{SEFD}_{XX}^2+\text{SEFD}_{YY}^2}\nonumber \\
&=& \frac{1}{2}k\sqrt{\left(\frac{\tau_X T_{sysX}}{A_{rX}}\right)^2 + \left(\frac{\tau_Y T_{sysY}}{A_{rY}}\right)^2 }
\label{eqn:SEFD_rms_proof}
\end{eqnarray}
and
\begin{eqnarray}
A_r &=& \frac{\eta_0}{2}\frac{\Re\{Z_{LNA}\}}{\left| Z_{LNA} \right|^2}\norm{\mathbf{l}}^2.
\label{eqn:Ar2L}
\end{eqnarray}
where $A_r$ can be interpreted as the area of the array such that for a given incident plane wave with power density $P_{inc}$, the power delivered to the load is $P_{load} = A_rP_{inc}$. Alternatively, it can also be expressed as $A_r = \tau A_e$.

As demonstrated in Appendix~\ref{apendx:links}, the quantity $\tau$ relates to $R$. Substituting Eq.~\eqref{eqn:Ar2L} and Eq.~\eqref{eqn:Rx_tau} into \eqref{eqn:SEFD_rms_proof} and simplifying yields,
\begin{eqnarray}
\text{SEFD}_I^{rms}&=&\frac{4 k}{\eta_0}\sqrt{\frac{T_{sysX}^2R_X^2}{\norm{\mathbf{l}_X}^4}+\frac{T_{sysY}^2R_Y^2}{\norm{\mathbf{l}_Y}^4}}
\end{eqnarray}

\onecolumn
\section{Simulated and observed $\text{SEFD}_{XX}$ and $\text{SEFD}_{YY}$}
\label{apendx:XXYY_comp}
\begin{figure}[htb!]
\centering
          \begin{subfigure}{.45\textwidth}
      \includegraphics[width=\textwidth]{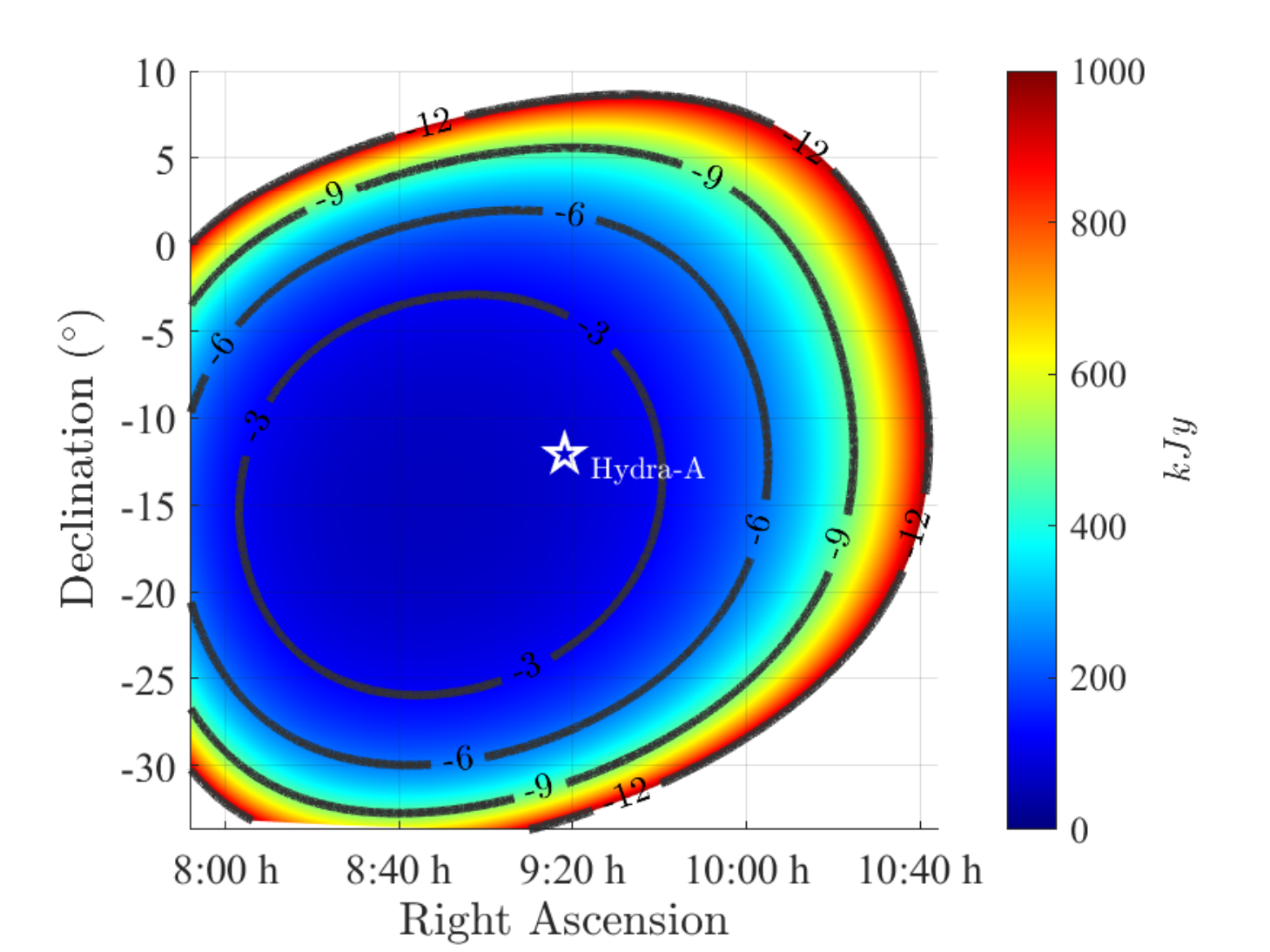}
      \caption{Simulated $SEFD_{XX}$}
      \label{fig:sim_sefd_xx}
      \end{subfigure}
        \begin{subfigure}{.45\textwidth}
     \includegraphics[width=\textwidth]{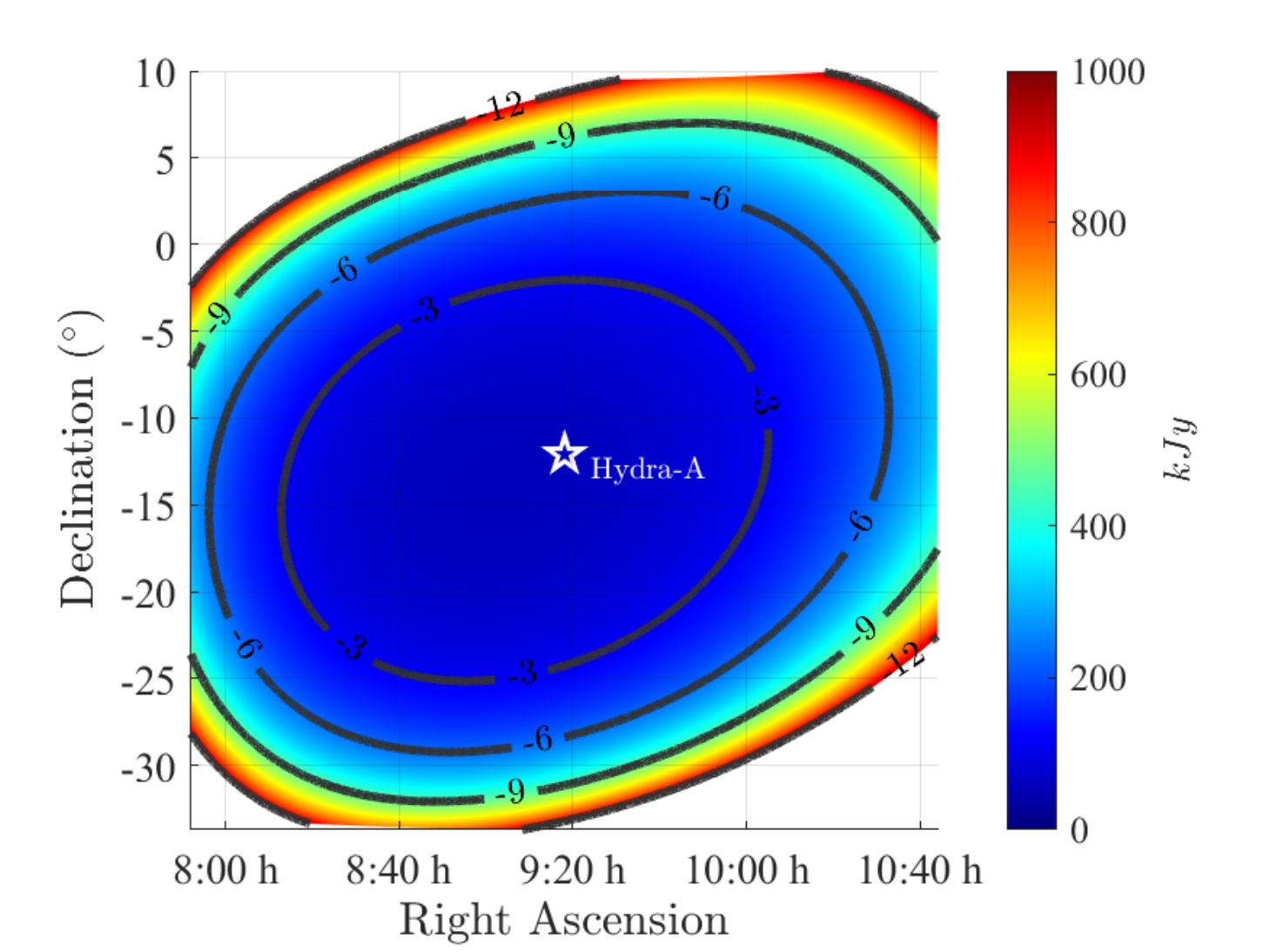}
      \caption{Simulated $SEFD_{YY}$}
      \label{fig:sim_sefd_yy}
      \end{subfigure}
            \begin{subfigure}{.45\textwidth}
     \includegraphics[width=\textwidth]{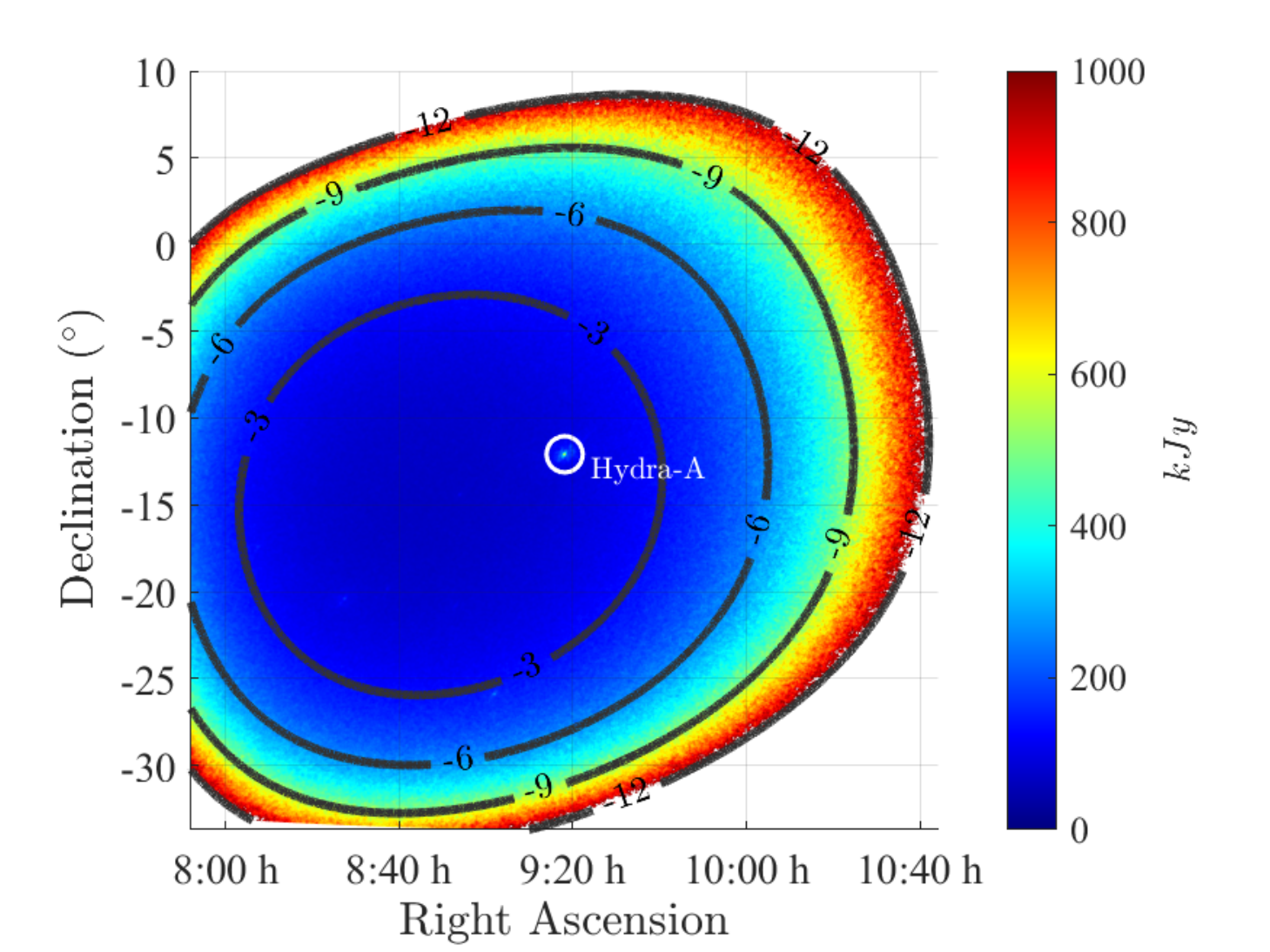}
      \caption{Observed $SEFD_{XX}$}
      \label{fig:obs_sefd_xx}
      \end{subfigure}
              \begin{subfigure}{.45\textwidth}
      \includegraphics[width=\textwidth]{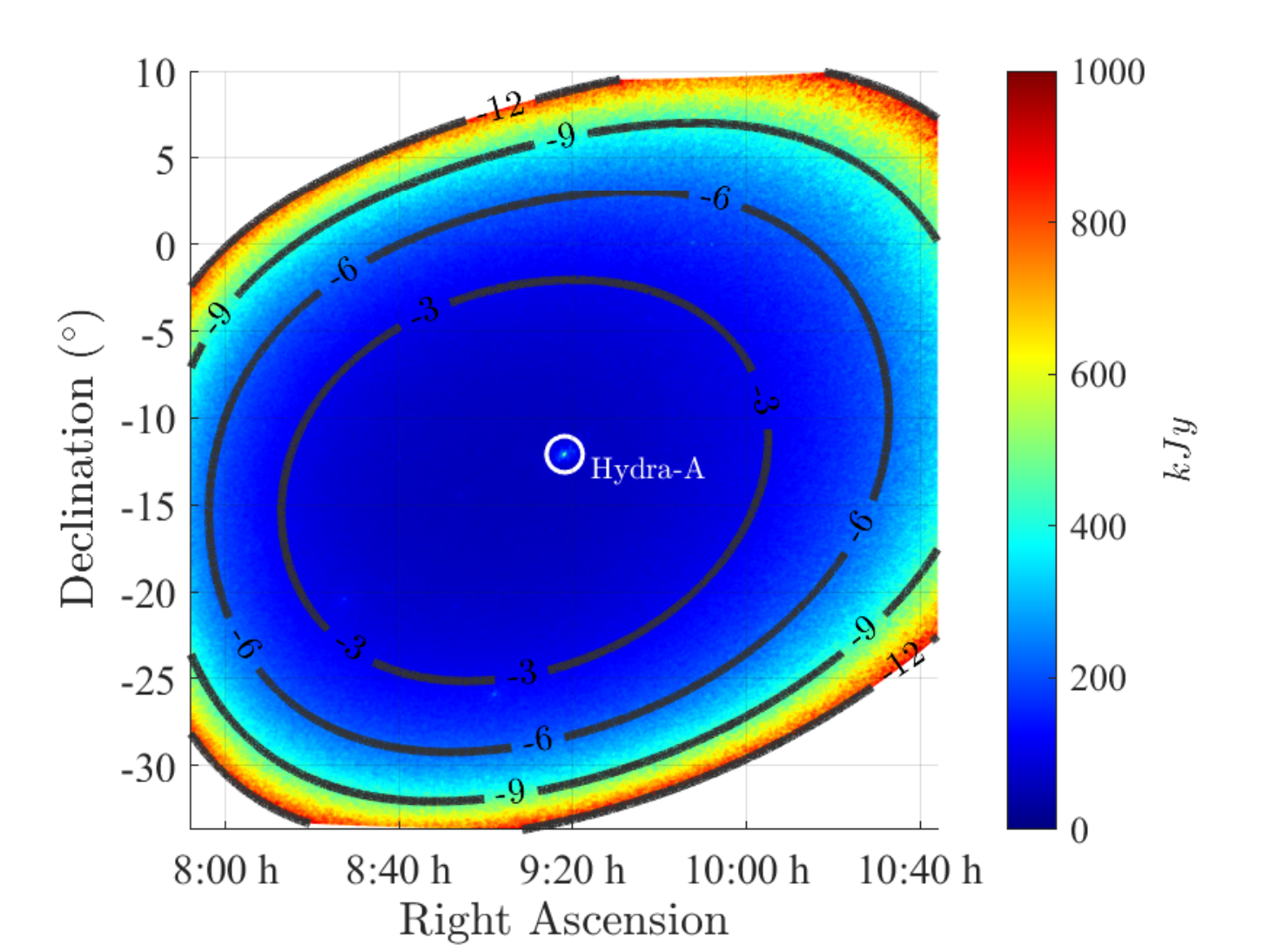}
      \caption{Observed $SEFD_{YY}$}
      \label{fig:obs_sefd_yy}
      \end{subfigure}
      \begin{subfigure}{.45\textwidth}
      \includegraphics[width=\textwidth]{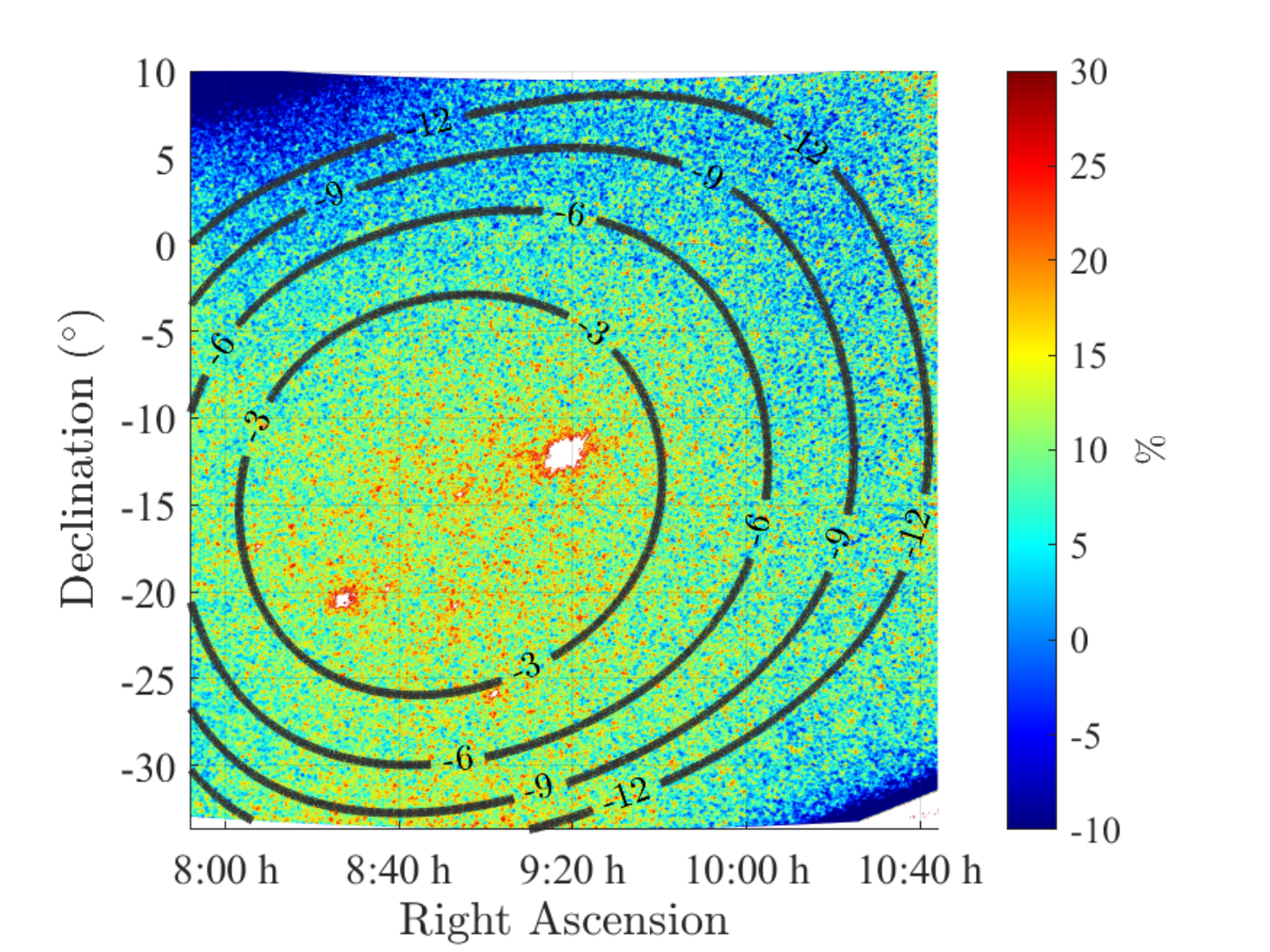}
      \caption{$M_{\text{SEFD}_{XX}}$, \%}
      \label{fig:delta_sefd_xx}
      \end{subfigure}
       \begin{subfigure}{.45\textwidth}
      \includegraphics[width=\textwidth]{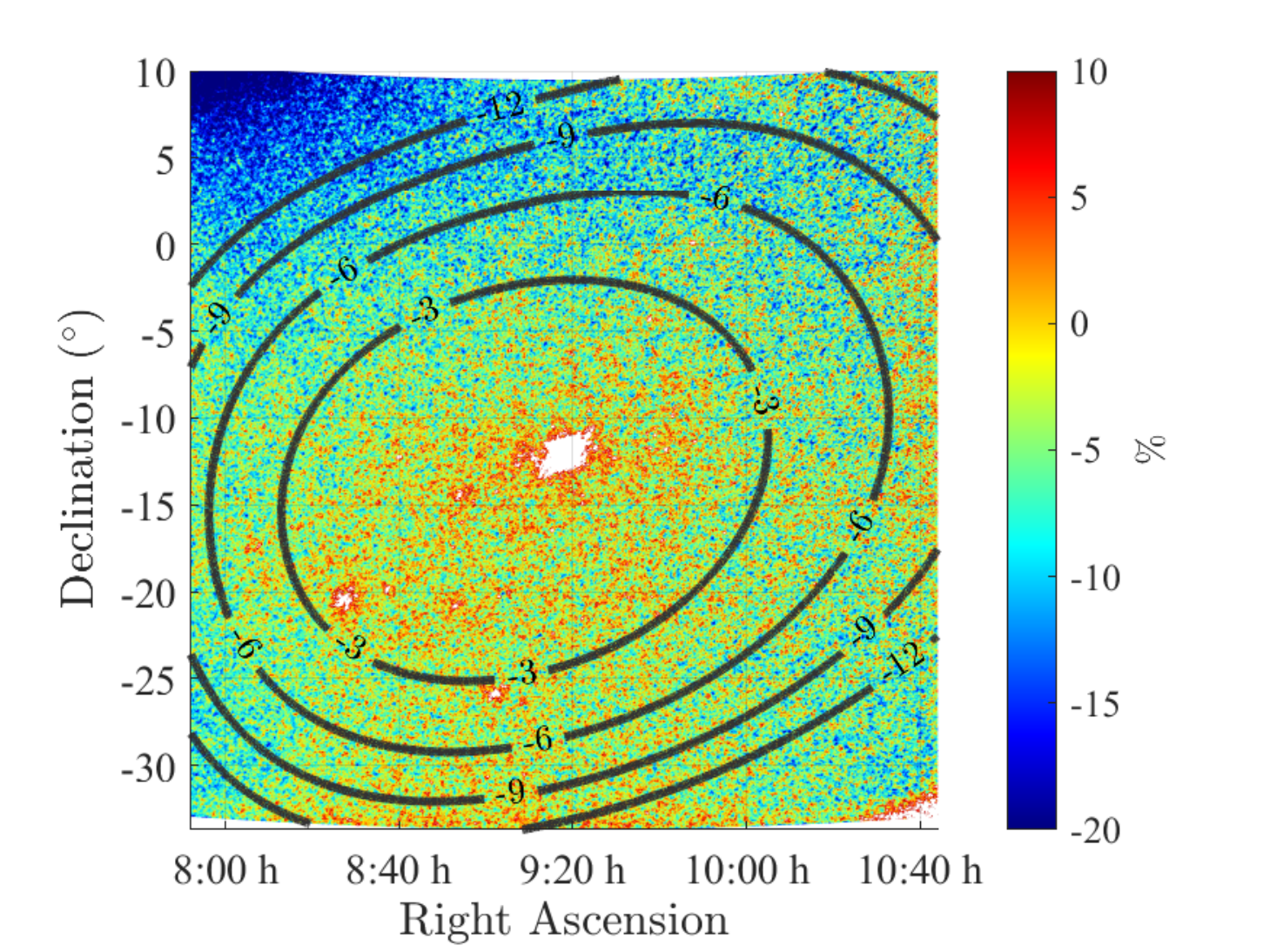}
      \caption{$M_{\text{SEFD}_{YY}}$, \%}
      \label{fig:delta_sefd_yy}
      \end{subfigure}
      \caption{Comparison of simulated and measured $\text{SEFD}_{XX}$ and $\text{SEFD}_{YY}$. Relative difference $M_{\text{SEFD}}$, \% is evaluated for each pixel.} 
      \label{fig:sefd_sim_meas_compar_XXYY}
\end{figure}

\end{appendix}

\end{document}